\documentclass[format=acmsmall, screen=true, review=false,nonacm]{acmart}
\usepackage{booktabs}
\usepackage{xcolor}
\usepackage{float}
\usepackage[caption = false]{subfig}
\usepackage{mathtools}
\usepackage{enumitem}

\graphicspath{{./figures/}}

\makeatletter
\newenvironment{algo}{\begin{figure}[H]}{\end{figure}}

\newcommand{\algtrue}{\textbf{true}}
\newcommand{\fxn}[1]{\textup{\texttt{#1}}}
\newcommand{\arry}{\mathbf}
\newcommand{\oc}[1]{\text{\upshape\texttt{#1}}}
\newcommand{\st}[1]{\mathcal{#1}}
\newcommand{\qpoint}{\oc{qpoint}}

\newcommand{\pforest}{\texttt{\upshape{p4est}}}

\newcommand{\cons}{consumer}
\newcommand{\pro}{producer}
\newcommand{\conscolorname}{orange}
\newcommand{\procolorname}{blue}
\newcommand{\Ppro}{P}
\newcommand{\Pcons}{Q}
\newcommand{\figref}[1]{Figure~\ref{#1}}
\newcommand{\secref}[1]{Section~\ref{#1}}
\newcommand{\algref}[2]{Algorithm~\hyperlink{#1}{#2}}
\newcommand{\appref}[1]{Appendix~\ref{#1}}
\newcommand{\worddef}[1]{\textit{#1}}

\acmJournal{TOMS}
\title[Overset computation of parallel distributed meshes]
      {Scalable overset computation between\\
       a forest-of-octrees- and an arbitrary\\
       distributed parallel mesh}
\author{Hannes Brandt}
\affiliation{%
	\institution{Rheinische Friedrich-Wilhelms-Universit\"at Bonn}
	\city{Bonn}
  \country{Germany}}
  \email{brandt@ins.uni-bonn.de}
\author{Carsten Burstedde}
\affiliation{%
	\institution{Rheinische Friedrich-Wilhelms-Universit\"at Bonn}
  \city{Bonn}
  \country{Germany}}
  \email{burstedde@ins.uni-bonn.de}
\thanks{%
The authors gratefully acknowledge partial support under DARPA Cooperative
Agreement HR00112120003 via a subcontract with Embry-Riddle Aeronautical University.
This work is approved for public release; distribution is unlimited.
The information in this document does not necessarily reflect the position or the
policy of the US Government.

We acknowledge additional funding by the Bonn International Graduate School
for Mathematics~(BIGS) as a part of the Hausdorff Center for Mathematics~(HCM)
at the University of Bonn.
The HCM is funded by the German Research Foundation~(DFG) under Germany’s excellence
initiative EXC 59 – 241002279 (``Mathematics: Foundations, Models, Applications'').
We acknowledge further funding by the DFG under grant no.\ 467255783
(``Hybride AMR-Simulationen'').

We would also like to acknowledge access to the ``Bonna'' and ``Marvin''
compute clusters hosted by the University of Bonn.%
}
\begin{document}

\ccsdesc[500]{Theory of computation~Sorting and searching}
\ccsdesc[500]{Theory of computation~Computational geometry}
\ccsdesc[100]{Mathematics of computing~Mathematical software performance}
\ccsdesc[500]{Computing methodologies~Massively parallel algorithms}
\ccsdesc[300]{Computing methodologies~Distributed algorithms}
\ccsdesc[500]{Computing methodologies~Distributed simulation}
\ccsdesc[100]{Computing methodologies~Massively parallel and high-performance simulations}
\ccsdesc[300]{Applied computing~Physics}

\keywords{Adaptive mesh refinement, forest of octrees, mesh overset, remote
search}

\begin{abstract}
	We introduce an algorithm that performs a one-directional mesh overset of a
	parallel forest of octrees with another distributed mesh of unrelated
	partition.
	The forest mesh consists of several adaptively refined octrees.  Individual
	smooth mappings for every tree allow to represent a broad range of geometric
	domains.
	The other mesh is generic and defines a distributed set of query points,
	e.g.\ stemming from a quadrature rule applied to each cell.
	We face the problem of finding data for all queries in the remote forest.
	
	The forest is partitioned according to its natural Morton ordering.
	Thus, the partition boundaries can be encoded globally with one Morton index 
	per
	process,
	which allows for precise, communication-free
	searching of the queries in the partition geometry.
	This is necessary to organize non-blocking communication of the queries to the
	relevant processes only.
	In a subsequent local search of the forest, we process the incoming queries
	and return the data of interest to each query's origin.
	
	The algorithm can be generalized, for example to
	load balancing of the overset, or adaptive refinement of the meshes around
	their intersection area.
	In 2D and 3D example scenarios we demonstrate the algorithm's performance
	and scalability to 12,288 processes.
\end{abstract}

\maketitle

\section{Introduction}
Many physical systems are modeled using multiple computational meshes.
A common example are Computational Fluid Dynamics (CFD) models around
multiple, sometimes deforming, objects, e.g.\ for simulations of wind
farms~\cite{KirbyBrazellYangEtAl19} or
rotorcraft~\cite{JarkowskiWoodgateBarakosEtAl13}.
A single mesh approach would require complicated geometric operations or
frequent re-meshing.
Instead, multiple meshes can be used in the Chimera
grid method~\cite{StegerDoughertyBenek83} (also known as mesh overset).

Next to the aforementioned geometrical considerations, a multi-mesh
simulation can also be motivated by physical properties of the system, e.g.\
when several physics layers interact.
For example, this occurs when modeling acoustic and gravity waves stemming from seismic sources~\cite{ZettergrenSnively19}.
The waves originate on earth and pass through the lower atmosphere as well as the ionosphere.
An efficient approach is to subdivide the domain into two overlapping parts
and to employ separate solvers for atmosphere~\cite{Snively13} and
ionosphere~\cite{ZettergrenSnively15}.
In particular, this allows the solvers to operate in different coordinate
systems to facilitate efficient solutions for each model.

To obtain physically meaningful results, the solvers of a multi-mesh
simulation must yield consistent solutions in their overlap regions.
Thus, they generally exchange simulation results, most commonly by first
searching and then interpolating solution data from one mesh to the other.

In the context of the Chimera grid method this process is referred to as donor
search~\cite{RogetSitaraman14,NoackBogerKunzEtAl09}.
This procedure is commonly accompanied by a hole profiling step, where cells
lying inside the solid body described by another mesh are identified.
Additionally, cells are assigned donor and receptor roles by choosing the
cell with the most accurate solution.

In this paper we consider mesh overset for applications like the geophysical
simulation described above.
There, hole profiling is not required as the domains do not define holes.
Furthermore, it is sufficient to pass interpolated data only from the
ionosphere model to the lower atmosphere model, resulting in a
inherently one-directional process without any need for cell-based role
assignment.
Consequently, in the scenario considered the overset consists of a donor
search combined with interpolation and communication of the results.

%

There is a variety of donor search algorithms for different mesh types.
For uniform, structured grids the donor cell can be computed directly.
For unstructured grids of various element types, more intricate solutions are
required.
Known strategies start with structuring the search space by storing it in a
coarse grid~\cite{RogetSitaraman14,HedayatAkbarzadehBorazjani22}, a K-d
tree~\cite{HorneMahesh19}, an Alternating Digital
Tree~\cite{RogetSitaraman14,SuhsRogersDietz02,NoackBogerKunzEtAl09,SchluterWuVanDerWeideEtAl05} or an octree~\cite{Gagnon22}.
After narrowing down the search space using such structures, a line
walk~\cite{RogetSitaraman14,NoackBogerKunzEtAl09}, a stencil
jump~\cite{SuhsRogersDietz02} or an exhaustive search~\cite{RogetSitaraman14}
can be applied to identify the actual donor.

Many mesh overset approaches employ distributed-memory parallelism to allow for
simulations on large scale clusters.
This introduces another obstacle to the donor search as now donor cells have to
be identified across processes.
Some approaches address this issue by replicating the query points on all
processes~\cite{Schwarz05} or by identifying target processes for the query
points based on axis-aligned bounding boxes of the mesh
partitions~\cite{HorneMahesh19}.
Other approaches distribute the domain such that overlapping meshes end up on
the same process~\cite{RogetSitaraman14,ZhaoLiGuoEtAl24,ShibilyevSezai21}.

This paper is concerned with the mesh overset of octree-based meshes.
Octrees are already commonly used in overset methods as structured,
axis-aligned (often called background)
meshes~\cite{PeronBenoit13,JudeSitaramanWissink21}, to organize and load
balance domain decomposition methods~\cite{ShibilyevSezai21}, for hole
profiling~\cite{NoackKadanthot02} or as a structure for process-local donor
searches~\cite{Gagnon22}.
However, their parallel search capabilities in combination with the geometrical
flexibility of a mapped forest-of-octrees structure have not yet been explored
fully in the overset context.

We consider a forest-of-octrees structure~\cite{BursteddeWilcoxGhattas11},
which only stores the leaves of the trees, thus avoiding the memory overhead of
maintaining several overlapping levels of cells and associated data.
For parallel distribution, a linear ordering of all leaf cells according to
the Morton space-filling curve~\cite{Morton66} is split into consecutive
sub-ranges of equal size, resulting in a partitioning fit for large scale
parallel applications.
By applying smooth mappings to the trees, a forest can represent a
broad range of geometric domains
\cite{BursteddeStadlerAlisicEtAl13}.
Hence, forests of octrees can serve not only as background mesh, but also as
adaptive, curvilinear 'near-body' meshes.

In the following chapters, we will introduce an efficient, scalable algorithm
performing mesh overset of a parallel distributed forest of octrees with an
unstructured mesh of unrelated partition, which consists of elements of
arbitrary type.
The unstructured mesh is represented by parallel distributed sets of
points, e.g.\ stemming from a quadrature rule applied to each cell, for which
evaluation data is queried from the forest.

The Morton-order based partitioning scheme allows for efficient donor searches
in the parallel distributed forest of octrees.
The exact partition boundaries of the mesh can be deduced from the first local
tree and Morton indices on every process.
We utilize this to efficiently search the queries in the forest
partition~\cite{Burstedde20d} and thereby assign each point to its donating
process.
Subsequently the points are searched in the local part of the forests employing
a top-down tree traversal for multiple query points at
once~\cite{IsaacBursteddeWilcoxEtAl15}.
By applying the inverse mapping of a tree to all query points -- similar
to~\cite{NoackBogerKunzEtAl09,HenshawSchwendeman08} -- we are able to avoid
expensive intersection tests with curved cells.

The back-and-forth messaging of the query points between the unstructured
mesh communicator and the forest communicator is based on non-blocking,
point-to-point communication.
Thus, computation and communication are overlapped and load imbalances are
partially compensated.
Additionally, the linear Morton ordering of the leaf cells allows for precise
and flexible repartitioning of the forest.
By assigning weights to each cell according to its overset workload and
performing a weighted repartitioning, the load balance of the overset can be
improved significantly.
The overset can be performed for congruent but also for disjoint or partially
overlapping mesh communicators.

The algorithm is defined in a modular manner by treating query points as
black box, user-defined structures, that are handled by user-defined
intersection tests and evaluation functions.
Thus, it can be repurposed with ease, e.g.\ for the weight computation of the
load balancing procedure or for adaptive refinement of both meshes around their
intersection area.

To summarize, we introduce an efficient, scalable, parallel distributed
one-directional mesh
overset algorithm for forests of octrees with an unstructured second mesh.
Tree-wise smooth mappings allow the forest mesh to cover curvilinear, adaptive
domains, as required for a near-body mesh.
The Morton-order based partitioning of the forest cells can be leveraged for
scalable donor searches in the distributed mesh and allows for easy yet
effective load balancing of the overset.
The proposed algorithm was implemented in the forest of octrees library
\pforest~\cite{Burstedde25} and tested for various artificial test
cases, showcasing near optimal scalability up to 12,288 cores.
It is applicable as is to simulations requiring a one-directional exchange,
e.g.\ in acoustic and gravity wave modeling.

\section{A Distributed Forest Data Structure}\label{sec:distributed_forests_of_octrees}

The focus of this paper is on mesh overset scenarios that include at least one distributed forest of octrees.
In this chapter we will give a minimal overview of a representer data
structure
and its key properties that our overset algorithm relies on.
For a comprehensive set of notation and definitions we refer to
\cite{IsaacBursteddeWilcoxEtAl15}.

First, we will present a general notion of forests of octrees and describe how the Morton order can be used to maintain a total order and distribute the octrees' cells~\cite{BursteddeWilcoxGhattas11}.
Then we will review search operations on octrees, both in a process local~\cite{IsaacBursteddeWilcoxEtAl15} as well as in a global manner~\cite{Burstedde20d}.
The ability to search parallel distributed points in a parallel distributed
forest of unrelated partition is a key element to our proposed algorithm.

\subsection{Distributed Forests of Octrees}\label{sec:octrees}

We define each octree on the cubic reference domain $[0,2^b]^3$.
Here, an octree is a tree with root $[0,2^b]^3$ and maximum level $b$ in
which each interior node has exactly eight children, which are derived by
splitting the parent node in half in every dimension.
The nodes of the tree are called \worddef{octants}.
The two-dimensional counterpart of an octree is a \worddef{quadtree}, which
divides each interior \worddef{quadrant} into four children.
All algorithms presented in this paper work with both octrees and quadtrees.

A \worddef{forest of octrees} can be seen as the conforming decomposition of
a geometric domain in physical space into $K \in \mathbb{N}$ subvolumes.
Each subvolume indexed by $k \in \mathbb{N}$ with $0 \le k < K$ is an octree
mapped from the reference domain into physical space by a diffeomorphism
$\psi_k \colon [0,2^b]^3 \to \mathbb{R}^3$.
The possibility to combine large amounts of trees with individual smooth mappings allows to represent a broad range of geometric domains,
of which a sphere or a torus (quadtrees mapped to a 2D surface in 3D) are relatively trivial examples.

In a linear forest of octrees \cite{SundarSampathBiros08} only the leaf
octants, which form a partition of the domain, are associated with data
(e.g.\ physical quantities).
The interior nodes are regarded as virtual objects that cover a cubical subtree of cells.
They are not stored at all, but rather created on the fly whenever needed.
Under this paradigm, distributing a forest of octrees comes down to
distributing its leaves.

The key to efficient partitioning of a forest of octrees is a total order of
the leaf octants induced by a space-filling curve.
We use a recursive level-wise lexicographical order known as the Morton order \cite{Morton66}.
By combining the Morton index $m$ of the front lower left corner coordinates of an octant with its level $\ell$ and tree index $k$ we obtain a total order over $(k,m,\ell)$.
Thereby, the task of distributing a forest of octrees to a set of $P$ processes can be reduced to the task of partitioning the linear ordering of all leaf octants into contiguous subranges.
This simplifies and accelerates load balancing tremendously.


To describe the global partition, every process $0 \le p < P$ shares $k_p$
and $m_p$ of its local octant with lowest global index as the tuple $(k_p,m_p,\ell_p)$.
The volume of the tree belonging to process $p$ is uniquely defined by $(k_p,m_p,b)$ and $(k_{p+1},m_{p+1},b)$, where we use the maximum-resolution level $b$ for simplicity.
By gathering the set
\begin{equation}
	\arry F \coloneqq \lbrace (k_p,m_p,b) \mid p = 0,\ldots,P-1 \rbrace \cup \lbrace (K,0,b) \rbrace
\end{equation}
of \worddef{global first positions} on every process, we can exactly determine how the reference and the geometric domain are distributed between the processes without any further communication~\cite{BursteddeWilcoxGhattas11, Burstedde20d}.
This is a key capability for remote searching in distributed forest of octrees.
\begin{figure}
	\centering
	\includegraphics[height=0.5\textwidth]{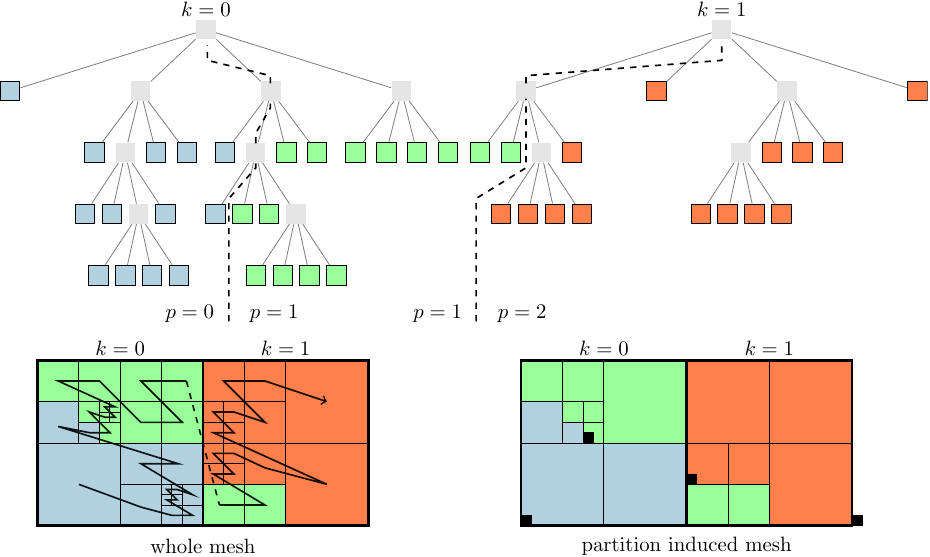}
	\caption{
		Example of a forest of $K=2$ quadtrees (top) and the corresponding mesh in the geometric domain (bottom left).
		The black arrow shows the ordering of the mesh quadrants according to the
		Morton space-filling curve.
                In this example the curves of both trees have the same orientation, but this is not necessarily so.
		The mesh is partitioned among three processes represented by colors and
		divided by black dashes.
		The mesh on the bottom right is the coarsest mesh suited to represent the partition boundaries exactly.
                It can be deduced communication-free from the global first positions
                (small, black cells) and is the basis of the partition
                search, even without ever being constructed explicitly.}
	\label{fig:octree_partition}
	\Description{
		Partition of a forest of octrees in tree and mesh representation.

		There are two trees in tree representation.
		Below them are two mesh representations of the forest of octrees, each
		consisting of squares of different sizes that combine to cover a two by one
		rectangle.
		The left mesh is labeled 'whole mesh'.
		The right mesh is labeled 'partition induced mesh'.

		The tree representation have different levels.
		The top level contains only one node per tree, the root.
		Each interior node is connected to exactly four nodes on the next level.
		The leaf nodes are all assigned one of three colors displaying the process
		they belong to.
		There are two dashed lines that intersect all interior nodes that have
		descendants of different colors and separate the leaves into three groups
		of distinct color.
		The squares of the whole mesh correspond to leaf nodes of the trees and are
		colored in the same three colors used before.
		The Morton space-filling curve is displayed as a line running through all
		square centers.
		It first runs through all squares of one color, before running through all
		squares of the next color.
		The route of the line can be compared to self-similar repetitions of the
		letter 'Z' on  different scales.
		The squares of the partition induces mesh correspond to leaf or interior
		nodes of the tree.
		The mesh contains small squares representing the global first positions.
		The squares are smaller than the smallest squares depicted in the mesh.
		The squares are located at the positions where the Morton space-filling
		curve transitions from one square to another square with different color.
		An additional global first position is located outside the two by one
		rectangle, touching its boundary.
		It is located at a potential starting point for a three by one rectangle
	  expansion.}
\end{figure}
An example of a partition and how it is encoded by the global first positions is displayed in \figref{fig:octree_partition}.

\subsection{Searching in Distributed Octrees}\label{sec:searching_octrees}
\begin{figure}
	\centering
	\includegraphics[height=0.4\textwidth]{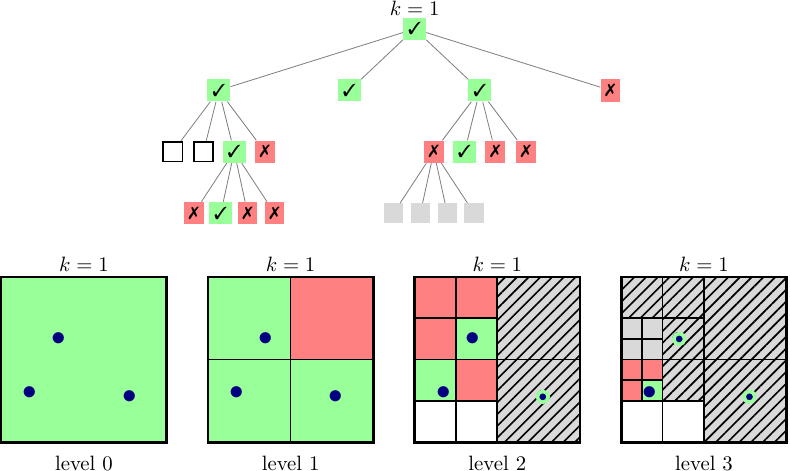}
	\caption{
		Local search of three points (blue) on process $2$ in the example forest from~\figref{fig:octree_partition}, shown for the tree $k_1$ (top) and the corresponding part of the mesh (bottom).
		Green cells containing a check mark indicate at least one point match in the given quadrant.
		Red cells containing a cross indicate no match.
		A gray cell indicates that the quadrant was not part of the search, either because it was visited before (hatched cells) or an ancestor cell yielded no match (plain cells).
		Cells in the tree belonging to a different process are white.
		The four meshes on the bottom describe the execution of the search by increasing levels of the tree.}
	\label{fig:octree_search}
	\Description{
		Local search of three points in a forest of octrees in tree and
		mesh representation.

		At the top, there is the tree representation of the second tree from
		the previous figure on octree partitions.
		Below, there are four mesh representations of the second tree from the same
		figure.
		From left to right they are labeled 'level 0' to 'level 3'.

		Some nodes of the tree representation contain checkmarks or crosses
		representing the flow of the top-down search.
		Checkmarks represent at least one point being found in the node.
		Crosses represent no point being found in the node.
		The root has a checkmark.
		If an interior node has a checkmark, all its children have checkmarks or
		crosses as well.
		If an interior node has a cross, all its children are marked grey, as they
		are not part of the search.
		There are two squares blanked out.
		They are non-local and will never appear in the search.
		They do not appear in the search despite a checkmark at their parent's node.
		The four mesh representations display the flow of the search on the
		different levels.
		For each level, the mesh contains only squares of this or a lower level.
		Lower level squares are hatched, as they are no longer part of the search.
		There are three query points.
		Whenever a square contains a query point, it is marked as success.
	}
\end{figure}

The mesh overset algorithm requires efficient searching of parallel distributed sets of points in the distributed forest of octrees.
Here, the distribution of the points may not match the distribution of the forest at all.
A communication-free \worddef{partition search}~\cite{Burstedde20d} can assign each local point to the process it intersects and thereby help organizing communication.
After all points are sent away and received by their respective processes, they can be searched process-local and assigned to a leaf octant~\cite{IsaacBursteddeWilcoxEtAl15}.

\subsubsection{Process-local Search}
\label{sec:localsearch}

Each process $p$ owns a contiguous subrange of the global ordering of the leaf octants with range $[\arry F_p, \arry F_{p+1}) = [(k_p,m_p,b),(k_{p+1},m_{p+1},b))$.
To search in the local part of the forest the trees in the range $[k_p,k_{p+1}]$ are traversed in a top-down manner for multiple points simultaneously, starting at the root with the complete set of local query points.
At each recursion the current octant gets tested against all
points that passed at its parent.
Only the points that yield a potential match get passed on to its children.

The root and all other interior octants are constructed temporarily on the fly.
Simultaneously, the contiguous subsection of the linear leaf array that intersects the current interior octant is tracked and updated~\cite{IsaacBursteddeWilcoxEtAl15}.
For octants that do not intersect the range $[\arry F_p, \arry F_{p+1})$ at all the recursion stops immediately.
An example of this \worddef{local search} can be seen in \figref{fig:octree_search}.

The search algorithm can be defined in a rather general manner by treating points as anonymous structures and relying on a user-defined callback function
\begin{equation}
	\texttt{Local\_Match}(\text{query point }\qpoint, \text{octant } \oc o,\text{flag } \texttt{isLeaf}).
\end{equation}
to decide which points intersect which octant.
The function has to return \texttt{true}, if $\qpoint$ may match with a descendant of $\oc o$ or $\oc o$ itself, and \texttt{false}, if this is certain to not be the case.
For interior nodes of the tree, where \texttt{isLeaf} is \texttt{false}, the callback may rely on fast, possibly false positive tests.
On the leaves, the callback is expected to check for exact matches.

This general approach allows for searching for a variety of objects, ranging
from points to rays, polygons, to any shape at all, legally intersecting
multiple cells at the same time.

\subsubsection{Partition Search}\label{sec:partition_search}
\label{sec:remotesearch}

The local search described in the previous section is sufficient to search a globally available set of points in a distributed forest of octrees.
However, if the points are distributed, e.g.\ because there are too many to store them on a single process or they originate from a distributed structure, a partition search is required.
This search routine operates on the global first positions $\arry F$ and assigns points (or any object rather) to the process(es) they are contained in.
The fundamental difference to the local search is that it has no access to, and thus does not reference, any leaves at all.

\begin{figure}
	\centering
	\includegraphics[height=0.5\textwidth]{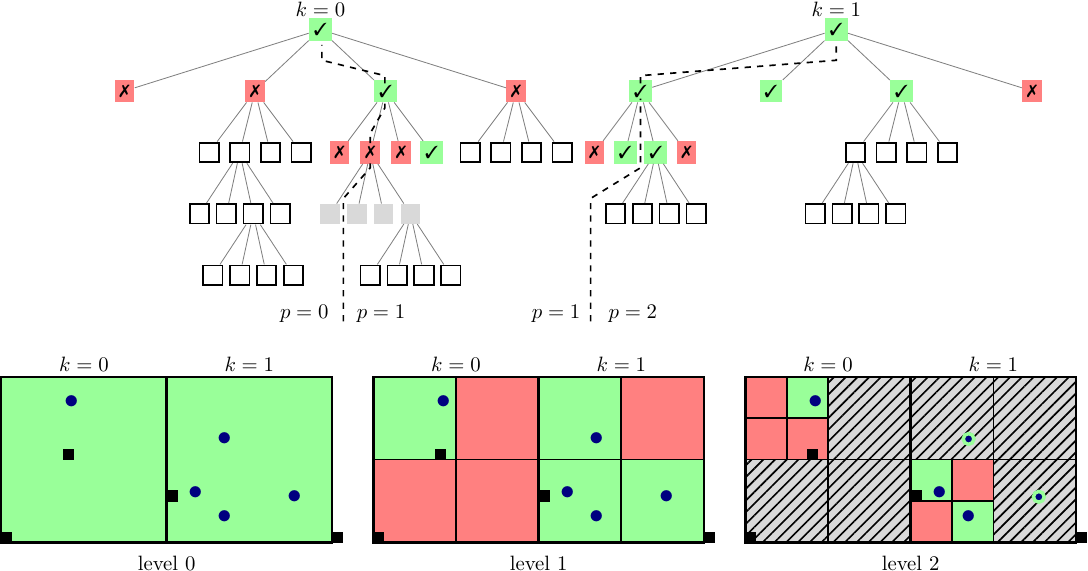}
	\caption{
		Partition search of five points (blue) in the example forest from~\figref{fig:octree_partition}, shown for both the trees (top) and the corresponding part of the mesh (bottom).
		Green cells containing a check mark indicate at least one point match in the given quadrant.
		Red cells containing a cross indicate no match.
		A gray cell indicates that the quadrant was not part of the search, either because it was visited before (hatched cells) or an ancestor cell yielded no match (plain cells).
		The global first positions (small, black cells) mark partition boundaries
		and thus trigger refinement of the search tree.
		Cells whose parent belongs to exactly one process cannot occur in the partition search and are marked in white.
		The search can be imagined to operate on the partition-induced tree from \figref{fig:octree_partition}.
		The three meshes on the bottom describe the behavior of the search on the different levels of the tree.
		Level $3$ was omitted since the search does not touch it.}
	\label{fig:octree_search_partition}
	\Description{
		Partition search of three points in a forest of octrees in tree and mesh
		representation.

		At the top, there is the tree representation of the forest from the figure
		on octree partitions.
		Below, there are three mesh representations of the forest from the same
		figure.
		From left to right they are labeled 'level 0' to 'level 2'.

		The nodes of the tree representation contain checkmarks or crosses
		representing the flow of the top-down search, similar to figure 2.
		Again, there are two dashed lines that represent the partition boundaries
		in the tree.
		If an interior node has a checkmark and is intersected by the partition
		boundaries, all its children have checkmarks or crosses as well.
		If an interior node is not intersected by the partition boundary, all its
		chidren are blanked out, as it will never appear in the search.
		The three mesh representations display the flow of the search on the
		different levels.
		For each level, the mesh contains only squares of this or a lower level.
		Lower level squares are hatched, as they are no longer part of the search.
		There are five query points.
		The global first positions are represented by small squares.
		Whenever a square contains a query point and a global first position, its
		children will be searched on the next level.
	}
\end{figure}

Similar to the local version the partition search traverses trees in a top-down manner searching multiple points simultaneously.
The difference lies in the parts of the forest that are traversed.
The partition search operates on all trees and stops the recursion as soon as an interior or leaf octant belongs to a single process $p$, so it is contained in the range $[\arry F_p, \arry F_{p+1})$.
Instead of a leaf octant with associated data it assigns just a process number to each point.
An example of the partition search can be seen in \figref{fig:octree_search_partition}.

Similar to the local search the algorithm relies on a callback
\begin{equation}
	\texttt{Partition\_Match}(\text{query point } \qpoint, \text{octant } \oc o,\text{rank } \texttt{pfirst}, \text{rank } \texttt{plast}).
\end{equation}
The parameters \texttt{pfirst} and \texttt{plast} that get passed to the \texttt{Match} callback describe the range of processes contributing to the current interior octant.
The search stops for $\texttt{pfirst} = \texttt{plast}$.
Thus, \texttt{pfirst} and \texttt{plast} have a similar role to the \texttt{isLeaf} flag in \texttt{Local\_Match}.
They are continuously updated during the forest traversal, splitting $\arry F$ into increasingly smaller index ranges.

Since the global first positions $\arry F$ are available globally and they
uniquely determine the partition boundaries, the partition search, like the
local search, operates communication-free.

\section{Overset Algorithm}\label{sec:overset_algorithm}

\begin{figure}
	\centering
	\includegraphics[width=.95\textwidth]{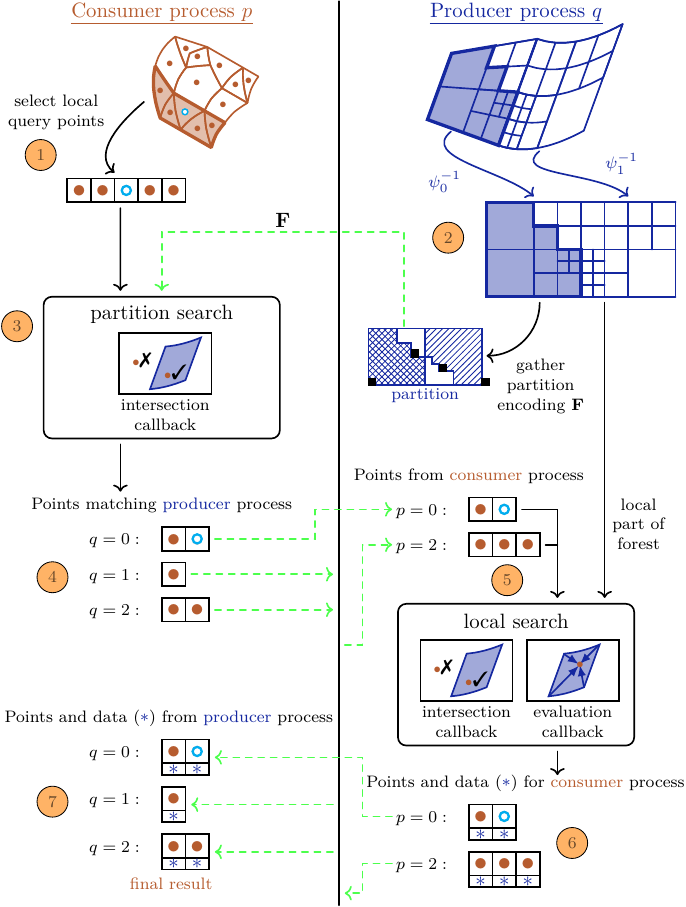}
	\caption{
		Flowchart of the mesh overset algorithm.
		The chart shows the local computations of rank $0$ of both the \cons{}
		(left, \conscolorname{}) and \pro{} (right, \procolorname{}) communicator
		as
		well
		as the communication they participate in (green, dashed lines).
		One query point is highlighted as a light blue circle during the whole algorithm.
		The numbers in orange circles offer a proposed reading order of the flow chart.
		The various steps are exposed in detail throughout \secref{sec:overset_algorithm}.
	}
	\label{fig:flowchart}
	\Description{
		Flowchart of the mesh overset for a process p on the consumer and a process
		q on the producer side.
		The whole flow of the algorithm is explained in detail in section 3.

		The chart is divided in a consumer-specific and a producer-specific half.
		Both sides show the respective mesh and how they deal with different query
		point buffers.
		At several steps, arrows indicate communication between the halves.

		There are numbers indicating the flow of the algorithm.
		In step 1, the centers of all five local cells in the consumer mesh - an
		unstructured combination of triangles, quadrilaterals and pentagons - are
		chosen as query points and stored in a linear array.
		In step 2, the producer processes gather the global first positions of the
		forest of octrees.
		In step 3, a partition search is performed.
		As input, it receives the query point array from step 1, the global first
		positions from step 2 and a intersection callback.
		As output, the search returns three arrays labeled 'q=0' to 'q=2' (step 4).
		Combined, they contain all five input query points.
		The query point arrays are communicated with the producer half, represented
		by multiple arrows.
		Some, but not all of the arrows arrive on the producer half of process p.
		In step 5, a local search is performed.
		As input, it receives the arrays received by process p, its local part of
		the forest and the intersection as well as a evaluation callback.
		As output, the search returns the input arrays together with an additional
		star representing interpolation data for each array entry (step 6).
		In step 7, various arrows indicate communication back to the consumer half.
		The query point arrays that resulted from step 3 now all have an additional
		star.
		This is marked as the final result.
	}
\end{figure}

In this section we proceed to detail all parts of the proposed parallel mesh
overset algorithm.
Its general flow is depicted in \figref{fig:flowchart}, and its key steps are
specified in the following sections.

First, we will describe the problem setup of the mesh overset, which
assumes representing general meshes by parallel distributed sets of
\worddef{query points}.
Then we will characterize how the partition and the local search from
\secref{sec:distributed_forests_of_octrees} can be used for searching
distributed query point sets in the forest-of-octrees mesh, before taking a
closer look at the communication between the two meshes.

\subsection{Problem Setup}\label{sec:problem_setup}
The proposed algorithm executes the overset of two meshes with at least one
of them, namely the mesh that is queried, being a distributed forest of
octrees.
It performs a one-directional mesh overset, where the roles of the two meshes are unambiguously distinguished,
and may serve as a building block for alternating or synchronous two-directional mesh overset algorithms.

A set of query points $\arry Q$ originating from a generic mesh is to be
located in a separate forest-of-octrees mesh to retrieve data.
How the query points are derived from the generic mesh is entirely up to the
user.
A common case would be to instantiate a quadrature rule in every one of its cells.
%
Strictly speaking, we would not even require any notion of mesh in the first
place but might think directly of a distributed parallel point cloud.
According to its role, we refer to this possibly unstructured mesh (or point
cloud) as the \worddef{\cons{}}.

We assume that each octant of the forest of octrees is associated with data
of interest, for example computed using a user-defined solver.
Our goal is to provide this data to the \cons{} by evaluation at its query
points, for example through interpolation of the solver's results.
Therefore, we refer to the forest of octrees as the \worddef{\pro{}}.
We do not make any assumptions on the nature of the interpolation -- our
algorithm is generic over any kind of information.

Both meshes may reside on congruent or on distinct sets of processors with different numbers of processes.
We denote the process count of the forest of octrees by $\Ppro$ and the process
count of the \cons{} mesh by $\Pcons$.
Consequently, we call a producer process $p$ and a consumer process $q$.
A parallel distributed \cons{} mesh leads to parallel distributed sets of query
points $\arry Q_q$ that form a partition of $\arry Q$.
So, the set of query points $\arry Q_q$ passed into the algorithm is a different one for every process $q$.

\begin{figure}
	\centering
	\includegraphics[width=0.9\textwidth]{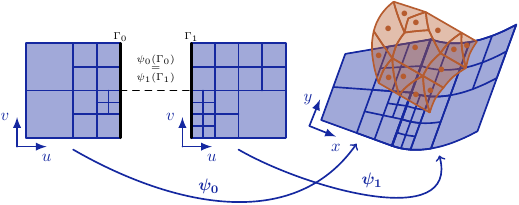}
	\caption{
		Example for the geometric setup of the mesh overset.
		The \cons{} side is displayed in \conscolorname{} and the \pro{} side in
		\procolorname{}.
		The \cons{} mesh is unstructured and consists of triangles, quadrilaterals
		and a pentagon.
		We have one query point for every cell center.
		The \pro{} mesh is a forest of two octrees.
                Both (square) reference domains are mapped to physical space
by a mapping $\psi_k$, respectively.
                The resulting geometric domain is connected, since the
trees intersect precisely at the edge $\psi_0 \left(\Gamma_1\right) = \psi_1
\left(\Gamma_2\right)$.
                Note that the parallel partition and distributed nature of
the two respective meshes is \emph{not} shown in this display.%
	}
	\label{fig:problem_setup}
	\Description{
		Two square reference octrees and their curved mapped version, which
		overlaps the unstructured consumer mesh.

		On the left, there are two octrees in axis-aligned mesh representation
		(like in the figure on octree partitions).
		Each has its own coordinate axes u and v.
		On the right, the same octrees are represented in their mapped version.
		Here, they share a common edge.
		Each reference octree is connected to its mapped counterpart by an
		individual mapping-arrow psi 0 respectively psi 1.
		The mapped forest of octrees intersects the unstructured consumer mesh from
		the flowchart.
		They share the same physical coordinate axes y and x.
	}
\end{figure}

Whether and where the \pro{} and \cons{} mesh intersect depends on how they are
mapped into physical space.
As described in \secref{sec:octrees}, the forest of octrees is transformed from the reference space into physical space by smooth functions $\psi_k$, with $k$ being the index of the tree.
This is just a convention, however, and we leave the definition of geometry entirely to the application.

How the \cons{} mesh is mapped into physical space only affects the creation
of the query points.
Namely, before passing the points to our algorithm, they have to be
converted into the same physical coordinate system that the forest of
octrees maps to.
In practice, this means that some knowledge of the producer's geometry
definition must be made available, hardcoded or communicated, to the
consumer.
Only when the producer's geometry metadata is distributed in parallel, we
may incur the need to define an intermediate reference coordinate system,
which we do not detail here for simplicity.
An example setup for the mesh overset is displayed in \figref{fig:problem_setup}.

We may summarize the objective as to provide every \cons{} process $q$ with
the evaluation result for all of $q$'s query points $\arry Q_q$ that
intersect any \pro{}'s domain in physical space.
To this end, we will specify two black-box operations that the user must
define and implement, (a) an intersection query of a point with a
hypothetical octant and (b) an evaluation query that returns the data for a
query point when it is found to be in a specific octant of the \pro{} mesh.
%

\subsection{Partition Search of Query Points}\label{sec:partition_search_of_query_points}

After the specification of the query points, each \cons{} process $q$ owns a
process-local set $\arry Q_q$
with coordinates in the \pro{} physical space.
However, the solution data required for evaluation resides on the \pro{} side.
Accordingly, the query points need to be sent to the correct \pro{} process.

Naively, this might be achieved by gathering the query points from all \cons{}
processes on all \pro{} processes.
For congruent mesh communicators, such would be achieved by an \texttt{Allgather} call.
For disjoint mesh communicators, each \cons{} process might send its points to
every \pro{} process via the global communicator resulting in a communication
scheme similar to an \texttt{Alltoall} between the communicators.

However, these schemes would result in predominantly unnecessary
communication, since each query point obtains its evaluation data from at
most one \pro{} octant and thereby at most one process.
Furthermore, they would exhaust the available memory already on moderately
sized parallel simulations.
Thus, we proceed in a decidedly tighter fashion.

\subsubsection{Proposed Approach}

Our method proceeds in two stages centered around a remote and a local mesh search.
With the information available on the \cons{} side we cannot determine the exact
octant, because we do not know the \pro{} forest's leaf structure.
Nonetheless, we can assign the point to the process that contains the octant
by use of the partition search described in \secref{sec:remotesearch}.
This only requires knowledge of the global first positions $\arry F$ of the
producer forest, which are cheaply communicated,
as well as the forest connectivity and geometric mappings, which can be
compiled into the consumer or otherwise shared.
We exemplarily describe the interchange for disjoint communicators in
\secref{sec:communication} and its implementation in
\appref{app:arc_disjoint_communicators}.
\begin{algo}
	\hypertarget{alg:partition_match}{}
	\includegraphics{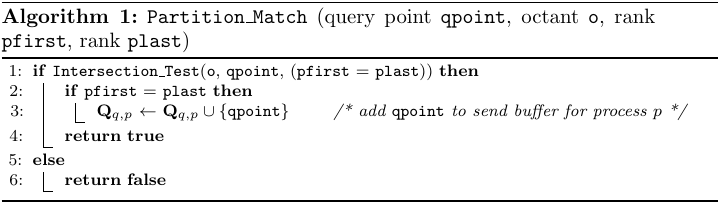}
\end{algo}

The partition search allows us to split the set $\arry Q_q$ into $\Ppro$
disjoint sets $\arry Q_{q,p} \subseteq \arry Q_q$ containing the query points
from \cons{} process~$q$ that require information from \pro{} process~$p$.
To accomplish this, we provide a callback function
$\fxn{Partition\_Match}$ (\algref{alg:partition_match}{1})
satisfying the call convention defined in \secref{sec:partition_search}.
Since we are searching for the donor of the query point $\qpoint$, we only return \texttt{true} if the octant $\oc o$ may contain $\qpoint$, which we delegate to a user-defined callback
\begin{equation}
	\texttt{Intersection\_Test}(\text{octant } \oc{o}, \text{query point } \qpoint, \text{boolean } \texttt{exact}).
\end{equation}
The intersection test must be provided by the application and may be
over-inclusive if $\oc{pfirst} \neq \oc{plast}$ (\texttt{exact} is
\texttt{false}), but it has to yield the precise result when there is only
one participating process left for the given $\qpoint$.
The reason is that we wish to to send $\qpoint$ only to the one process it
lies in, thus eliminating false positives.

During the partition search every process creates and constantly updates the
$\Ppro$ sets $\arry Q_{q,p}$ local to process $q$, which all start out empty.
Since we request an accurate intersection test when $\oc{pfirst} =
\oc{plast}$, every point will appear in at most one set if we apply a proper
tie breaking rule around octant boundaries.  The points that lie outside of
the \pro{} domain will not appear at all.

When the partition search is complete, $\arry Q_{q,p}$ contains all query
points that have to be sent to \pro{} process $p$ for evaluation.

\subsubsection{Intersection Test}\label{sec:intersection_test}
The callback in \algref{alg:partition_match}{1} depends on the mappings
$\psi_k$ applied to the forest of octrees as well as the representation of
the query points.

As mentioned in \secref{sec:problem_setup}, the query point $\qpoint$ is given
in the \pro{} physical coordinate system.
Thus, it is a straightforward approach to transform the octant $\oc o$ into the
same system by use of the mapping $\psi_k$ of the \pro{} tree containing $\oc
o$.
However, in some cases this can result in curved edges and faces of the
mapped octant which complicates the intersection test and might even require
solving a system of nonlinear equations.

\begin{figure}
	\centering
	\includegraphics[width=\textwidth]{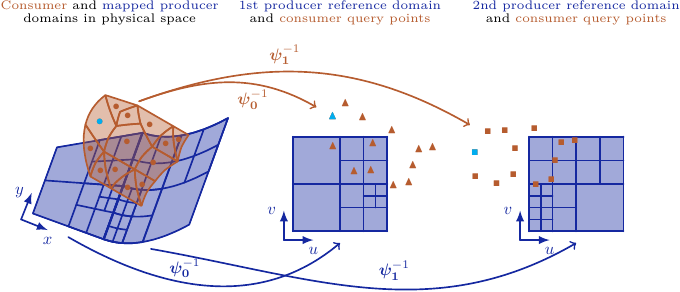}
	\caption{
		Example for the query point search in the reference domain.
		This continues the problem setup introduced in \figref{fig:problem_setup}.
		The \cons{} side is displayed in \conscolorname{} and the \pro{} side in
		\procolorname{}.
		The query points originate in the \cons{} mesh.
		They are mapped to the reference domain by applying $\psi_k^{-1}$ for every
		\pro{} tree $k$.
		The points for tree $1$ are represented by $\blacktriangle$ and the points for tree $2$ by \scalebox{0.8}{$\blacksquare$}.
		The resulting point clouds differ due to the different inverse mappings.
                Should the inverse map of points outside the reference image be undefined,
                we would expect the user to detect and discard these points ahead of time.
	}
	\label{fig:inverse_search}
	\Description{
		A forest of octrees overlapping a consumer mesh with query points;
		shown both in physical space and the reference space.

		On the left there is the mapped forest of octrees and the unstructured
		consumer mesh, both as described in the previous figure introducing the
		problem setup and labeled 'consumer and mapped producer domains in physical
		space'.
		On the right, there are the two octrees in axis-aligned mesh representation
		labeled '1st/2nd producer reference domain and consumer query points'.
		For each axis-aligned mesh there also is a separate clouds of consumer
		query points, overlapping it partially.
		The mapped trees are connected with the corresponding axis-aligned mesh
		representation by arrows labeled 'inverse of psi 0/1'.
		The consumer points from the unstructured mesh are connected to the two
		point clouds by arrows with the same labels.
  }
\end{figure}

To avoid this, in many cases we can inversely map $\qpoint$ into the reference space of $o$ instead, which is a common approach for donor searches in curvilinear grids~\cite{Meakin91,NoackBogerKunzEtAl09,HenshawSchwendeman08}.
This technique can be applied to mapped forests of octrees as well.
The key advantage of this approach is that it allows for easy and accurate intersection tests since all octant faces are planar and axis-aligned in the reference domain.
Especially, it comes with negligible additional cost to perform accurate tests at every level.
Due to the efficiency of the axis-aligned tests, computing $\smash{\psi^{-1}_k}$ of $\qpoint$ may be the most time consuming part of the $\fxn{Intersection\_Test}$.
However, it has to be computed and stored only once each time the search enters
a new tree $k$.
An example of this approach is displayed in \figref{fig:inverse_search}.

While the inverse mapping approach can be beneficial in many cases, it has
to be handled with care.
For example, $\psi_k^{-1}$ may not be defined for query points outside of
the mapped producer domain.
In this case, a tree intersection test in the \pro{} physical coordinate
system preceding the actual tree recursion can help to ensure that we only
inversely map points inside the image of $\psi_k$.
If no explicit formula for $\psi_k^{-1}$ is available, a Newton iteration
may be employed to approximate $\smash{\psi_k^{-1}}$ for any point.

\subsubsection{Technical Intricacies}
When implementing the intersection test, potential floating-point errors
need to be considered and handled.
This is the case even for the axis-aligned test based on inverse mappings:
Octant boundaries are in $([0,2^b] \cap \mathbb{Z})^3$ and can be
represented exactly either as an \texttt{integer} or as a \texttt{double}.
However, the application-dependent inverse mapping of the query point may still
be subject to floating-point errors.
This may result in a query point lying on an octant boundary inside the mesh
not being found by any of the adjacent octants.

To avoid missing any point inside the mesh, we recommend a tolerance to the
intersection test.
Thus, neighboring octants `overlap' around their shared faces, edges or corners.
This approach causes new inaccuracies in the search procedure, as a query point
may now be found in several octants.
However, this can be dealt with by removing the point from the search once it
was found in any octant.
Under the assumption of the evaluation function being smooth, erroneously
assigning a point to a direct neighbor octant causes only a moderate error.

If the tolerance leads us to identify a direct neighbor process by a small
margin, we must make sure that the ensuing process-local search, described
below, is guaranteed to match within that tolerance to not lose the point.

\subsection{Local Search and Evaluation}\label{sec:local_search}

After the partition search, the query points are sent from their respective
\cons{} processes to the \pro{} process that contains the data for them.
How to determine the communication pattern will be elaborated upon further
in \secref{sec:communication}.
For now, we assume that a given \pro{} process $p$ has received $\arry
Q_{q,p}$ from \cons{} process $q$.

Due to the intersection tests on the leaves of the partition search tree being
accurate, all incoming points are contained in the local part of the \pro{}
forest of octrees.
We can apply the local search routine described
in~\secref{sec:localsearch} to all received points simultaneously.
\begin{algo}
	\hypertarget{alg:local_match}{}
	\includegraphics{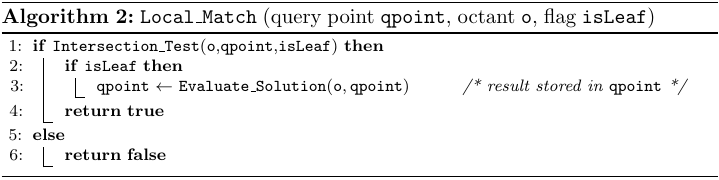}
\end{algo}

Similar to the partition search, we define a \texttt{Local\_Match} callback
function according to the convention of the search routines described
in~\ref{sec:searching_octrees}.
It accepts the \texttt{isLeaf} flag in lieu of the \texttt{pfirst}
and \texttt{plast} arguments.
The function may reuse the inverse intersection test from
\secref{sec:partition_search_of_query_points}.
Again, the test may yield over-inclusive results except when on the leaves
of the search tree, which is indicated by the \texttt{isLeaf} flag.

If a query point intersects a leaf octant, we can start with the actual evaluation of the solution data for the point, which is performed by a user-defined callback function
\begin{equation}
	\fxn{Evaluate\_Solution}(\text{octant } \oc{o}, \text{query point } \qpoint).
\end{equation}
Since the leaf octant is part of the local forest, we can assume that the
solution data is available on this process, which was generally not the case
during the partition search.
Finally, we store the resulting evaluation data in the query point structure it belongs to.
Afterwards, we send the resulting, data-augmented query point buffer
$\arry Q_{q,p}^*$ back to \cons{} process $q$.

\subsubsection{Technical Intricacies}
Similar to the partition search, floating-point errors need to be considered
when it comes to implementing the intersection test in the local search.
Again, adding a tolerance to the axis-aligned intersection test is a sensible
approach.

The subdivision of the global search routine into partition and local search
introduces another potential error source to the system.
Based on the results of the partition search the \cons{} process $q$ should
receive evaluation data from \pro{} process $p$ for \emph{every} query point in
$\arry Q_{q,p}$.
It is desirable to preserve this property despite potential floating-point
errors.

Thus, we have to make sure every query point found in the partition search is
found in the local search, too.
To achieve this, we relax the tolerance in the local search by a
factor of two in relation to the partition search.

The partition search compares the query points to the shallower octants
including roots, while the local search compares them to the deeper octants
including leaves.
Thus, the tolerances may not be chosen relative to the octant's size.
Instead, the tolerance should be a constant
below the minimum octant
size, e.g.\ $t = 1000 \varepsilon$ for the partition and $2 t$ for the local
search with machine $\varepsilon$.

\subsection{Communication}\label{sec:communication}
During the mesh overset all query point buffers $\arry Q_{q,p}$ are
sent from \cons{} process $q$ to \pro{} process $p$.
Eventually, the resulting updated buffers $\arry Q_{q,p}^*$ have to be returned.
Since we allow overlapping or even congruent \pro{} and \cons{} communicators, a
process might assume both roles at the same time.
Therefore, we need to intertwine \cons{} and \pro{} communication steps in a way
that avoids any potential deadlocks and race conditions.

The pseudocode of the resulting algorithm is displayed in
\algref{alg:overset}{3}.
There, unlike in the rest of this paper, $p$ denotes not a \pro{}, but a global
rank which may be a \cons{}, a \pro{} or both.
We use background colors to distinguish between the different roles $p$ may
assume.
The orange lines are only executed when $p$ is a \cons{} rank, the blue lines
are only executed when $p$ is a \pro{} rank, and the neutral lines are executed
when $p$ is both.
The inputs $\arry Q_p$ and $f_p$ are considered to be $\emptyset$ or
\texttt{NULL}, respectively, if $p$ does not have the corresponding role.

\subsubsection{Partition Search and Notification}
If the \cons{} and \pro{} mesh reside on congruent communicators, the partition
search operates communication-free \cite{Burstedde20d}.
%
For disjoint communicators we cannot avoid communication completely, because we
need to send the $\Ppro + 1$ global first positions $\arry F$, which encode the
partition of the forest of octrees, from the \pro{} to the \cons{} side.
To achieve this, rank $0$ of the \pro{} communicator sends $\arry F$ to rank
$0$
of the \cons{} communicator.
Then the \cons{} rank $0$ broadcasts the information to all processes in its
communicator.
This scheme may be optimized further, but there is usually no need.

\begin{algo}
	\hypertarget{alg:overset}{}
	\includegraphics{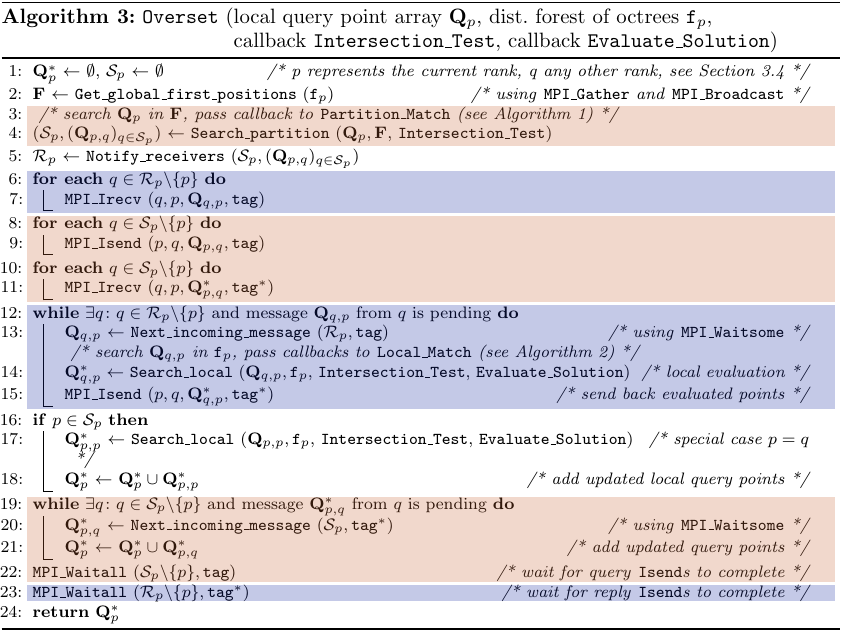}
\end{algo}
Depending on the application, the mappings $\psi_k$ relevant for the
intersection test may also change with time.
In this case, the two communicators additionally need to update according to
these changes in the same rhythm.

After the partition search, \cons{} process $q$ has built $\Ppro$ sets~$\arry
Q_{q,p}$.
Some of them might be empty, in which case no message needs to be sent at all.
We define
\begin{subequations}
\begin{align}
	\st S_q &\coloneqq \lbrace p \in \left[ 0,\Ppro \right) \cap \mathbb{Z} \mid \arry Q_{q,p} \neq \emptyset \rbrace \quad \text{for } q = 0,\ldots, \Pcons - 1,  \\
	\st R_p &\coloneqq \lbrace q \in \left[ 0,\Pcons \right) \cap \mathbb{Z} \mid \arry Q_{q,p} \neq \emptyset \rbrace \quad \text{for } p = 0,\ldots, \Ppro - 1.
\end{align}
\end{subequations}
The set $\st S_q$ contains all processes \cons{} process $q$ sends messages to
while $\st R_p$ contains all processes \pro{} process $p$ receives messages
from.
However, after the partition search $\st R_p$ is not yet known to \pro{} process
$p$.
Before the actual exchange of query points can begin, the \pro{} processes need
to be notified about their future communication partners $\st R_p$ and the
corresponding incoming message sizes $\lvert \arry Q_{q,p} \rvert$.
To achieve this, we employ a tree-based communication
scheme~\cite{IsaacBursteddeGhattas12, Burstedde20d}.

\subsubsection{Communication of Query Point Sets}
After the notification step the actual communication of the query points begins.
The communication system may be sparse and irregular:
Especially for locality-preserving partitioning schemes -- like the Morton
order based partitioning employed for the forest of octrees -- the process-local
part of the mesh covers only a small area of the domain.
If similar properties hold for the other mesh's partitioning, this small area
only intersects cells owned by a limited amount of processes.
Thus, \mbox{$\lvert \st S_q \rvert \ll \Ppro$} and \mbox{$\lvert \st R_p \rvert \ll \Pcons$} is a likely scenario.
Consequently, we avoid collective communication whenever possible.
Instead, we focus on non-blocking, point-to-point communication using \texttt{MPI\_Isend} and \texttt{MPI\_Irecv}.

First, every \pro{} process $p$ posts a non-blocking receive for all messages
$\arry Q_{q,p}$ with $q \in \st R_p$.
Next, every \cons{} processes $q$ starts non-blocking sends of these messages.
Since process $q$ knows it will eventually receive the updated query point buffers in return, it posts \texttt{MPI\_Irecv} calls for all $\arry Q_{q,p}^*$ with $p \in \st S_q$ right after.

On the \pro{} side, the actual evaluation process starts now.
The \pro{} processes wait for incoming messages using \texttt{MPI\_Waitsome}.
Whenever a message $\arry Q_{q,p}$ arrives, the process $p$ directly starts a local search of the point set in the local mesh, as it is described in \ref{sec:local_search}.
We start a separate call to the local search for every $\arry Q_{q,p}$.
Thereby, we miss out on some synergy effects, as we need to traverse the search tree $\lvert \st R_p \rvert$ times instead of once, albeit with proportionally less data.
However, this minimizes the idle time of \pro{} processes, as they can work on
every incoming message immediately.
As soon as the local search of $\arry Q_{q,p}$ is complete, the \pro{} process
returns the updated set $\arry Q_{q,p}^*$ to the \cons{} process it originated
from using \texttt{MPI\_Isend}.

The \cons{} processes wait for the arrival of the $\arry Q_{q,p}^*$, similar to
the \pro{} processes, using \texttt{MPI\_Waitsome}.
Whenever messages arrive, they are incorporated into the local query point set $\arry Q_q^*$ immediately.
A quadrature calculation can be executed on the cells for which all evaluation data
has already arrived.
Finally, all \cons{} and \pro{} processes wait for their non-blocking sends to
complete by a call to \texttt{MPI\_Waitall}.

If the \cons{} and \pro{} communicator overlap, one process may need to send
points to itself.
Instead, the process can directly access the \cons{} query point set $\arry
Q_{q,q}$ for its \pro{} local search.
Thereby, the evaluation of $\arry Q_{q,q}$ can be achieved without any communication and executed any time after the sets creation in the partition search and before the end of our algorithm.
We place it directly after the local search of all incoming messages sent from
\cons{} processes.
Thus, we do not stall the sets $\arry Q_{q,p}*$ from being returned to their
\cons{} side origin.
At the same time, we cover potential load imbalances of the previous steps,
which can lead to \cons{} processes idling while waiting for the incoming
updated query points.

\section{Numerical Applications}\label{sec:applications}
In order to test the scalability of the algorithm introduced in the previous
Section, we apply it to several artificial examples.
We start with a two-dimensional example of two perfectly overlapping meshes
based on simplistic mappings.
To showcase the algorithm's capability to deal with highly adaptive meshes, we
refine both of them from level $3$ to $20$.
Next, we increase the problems' complexity by switching to a three-dimensional
mapped setting.
Here, as an example of a derived novel functionality, we employ the overset
algorithm not only for interpolation, but also for guiding the adaptive
refinement of the meshes around their intersection area.

Our implementation is realized using the
\pforest~\cite{BursteddeWilcoxGhattas11} software library for parallel
adaptive mesh refinement on forests of octrees.
\pforest{} supplies the octree functionality necessary for our overset
algorithm: parallel partition along the Morton
curve~\cite{BursteddeWilcoxGhattas11}, mapping of the
trees~\cite{BursteddeWilcoxGhattas11}, and partition
search~\cite{Burstedde20d} as well as local search
routines~\cite{IsaacBursteddeWilcoxEtAl15}.
All examples are partition independent, which is a key property to ensure
reproducibility of the same problem across varying parallel environments.
The implementation can be found in the \texttt{example/multi} directory of the
\pforest{} repository~\cite{BursteddeWilcoxIsaac25} under the commit
\texttt{b34b7162}.

We perform strong and weak scaling tests on the on the Marvin MPP cluster for
up to 12,288 processes distributed over 128 compute nodes.
Each node contains two 48-core Intel Xeon ``Sapphire Rapids'' CPUs (2.10
GHz) and 1024 GB of memory.
To handle problem inherent load imbalances we introduce a weighted
repartitioning, which significantly improves the algorithm's scalability.

\subsection{Unit Square Example}\label{sec:unit_square_example}
As a first example we consider the 2D mesh overset of two forest-of-quadtree
meshes.
Both of them are refined adaptively from level 3 to level 20.
The \cons{} forest is designed solely for generating parallel
distributed sets of query points.

\subsubsection{Simulation Setup}\label{sec:unit_simulation_setup}

\begin{figure}
	\centering
	\includegraphics[width=\textwidth]{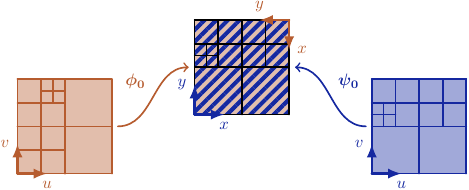}
	\caption{
		Geometric setup of the unit square example.
		The \cons{} side is displayed in \conscolorname{} and the \pro{} side in
		\procolorname{}.
    Both octrees are mapped to the same domain and refinement pattern in
    physical space, as indicated by the central, striped mesh and its two
    $(u, v) \mapsto (x, y)$ coordinate systems.
	}
	\label{fig:unit_problem_setup}
	\Description{
		A consumer and a producer unit square mesh in both axis-aligned mesh
		representation and in physical space.

		On the left, there is the consumer mesh in axis-aligned mesh representation.
		On the right, there is the producer mesh in axis-aligned mesh
		representation.
		Both consist of a single tree.
		In the middle, there are the mapped versions of the consumer and the
		producer trees in physical space, connected to their axis-aligned
		representation by an arrow labled phi 0 and psi 0, respectively.
		The mapped versions overlap completely, indicated by the mapped tree
		being striped in the consumer and producer colors.
		While the axis-aligned reference meshes are refined differently, their
		mapped refinement patterns match exactly.
	}
\end{figure}

\begin{figure}
	\centering
	\includegraphics[width=\textwidth]{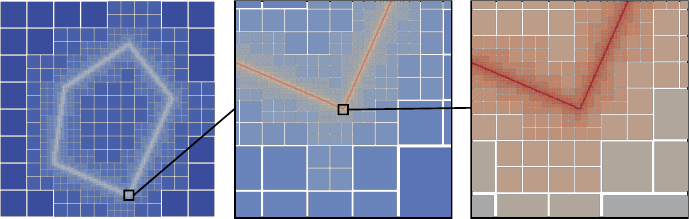}
	\caption{
		Levels of the adaptive unit square example.
		From left to right, the images subsequently zoom to a corner of the
		pentagon around which the mesh was refined, as indicated by the black boxes.
		The colors represent the different levels ranging from $3$ (blue) to $20$
		(red).
	}
	\label{fig:unit_levels}
	\Description{
		Three paraview-outputs of an adaptively refined, square forest of quadtrees
		showing its level of refinement with increasing zoom.

		On the left, there is an output of the entire mesh, with the level being
		color-coded.
		A pentagonal shape of high-level quadrants can clearly be recognized.
		The middle output shows a zoomed version of one of the corners of the
		pentagon.
		A clear cut line of quadrants of even higher refinement outline the corner
		of the pentagon.
		The middle output covers only a small part of the total mesh.
		The right output shows a zoomed version of a part of the middle output.
		The boundaries of individual quadrants of the highest level are still barely
		recognizable due to their small size.
	}
\end{figure}

\begin{figure}
	\centering
	\subfloat{\includegraphics[width=0.43\textwidth]{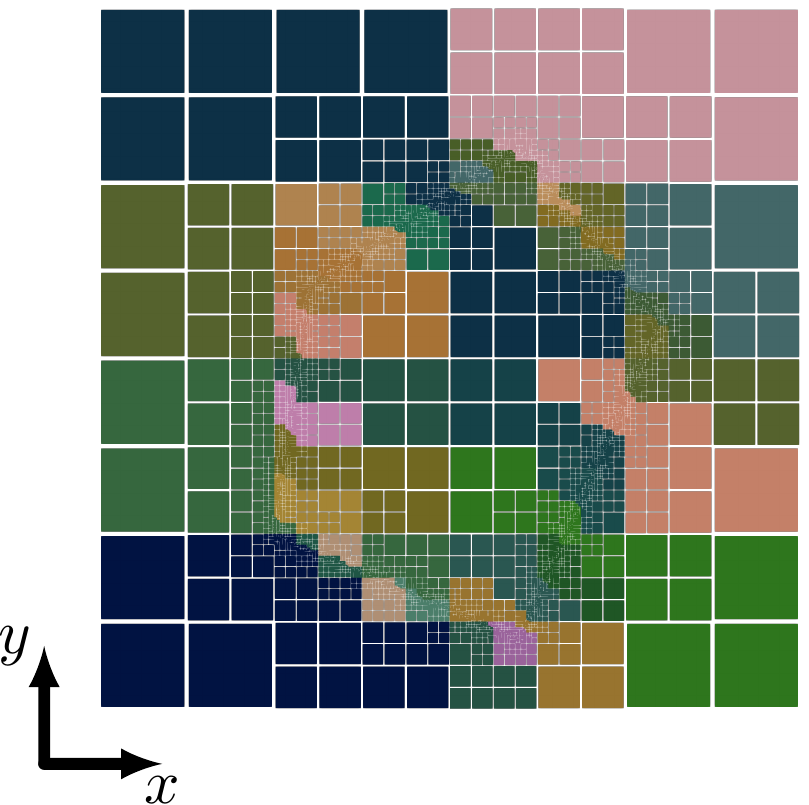}}
	\hspace{0.5cm}
	\subfloat{\includegraphics[width=0.43\textwidth]{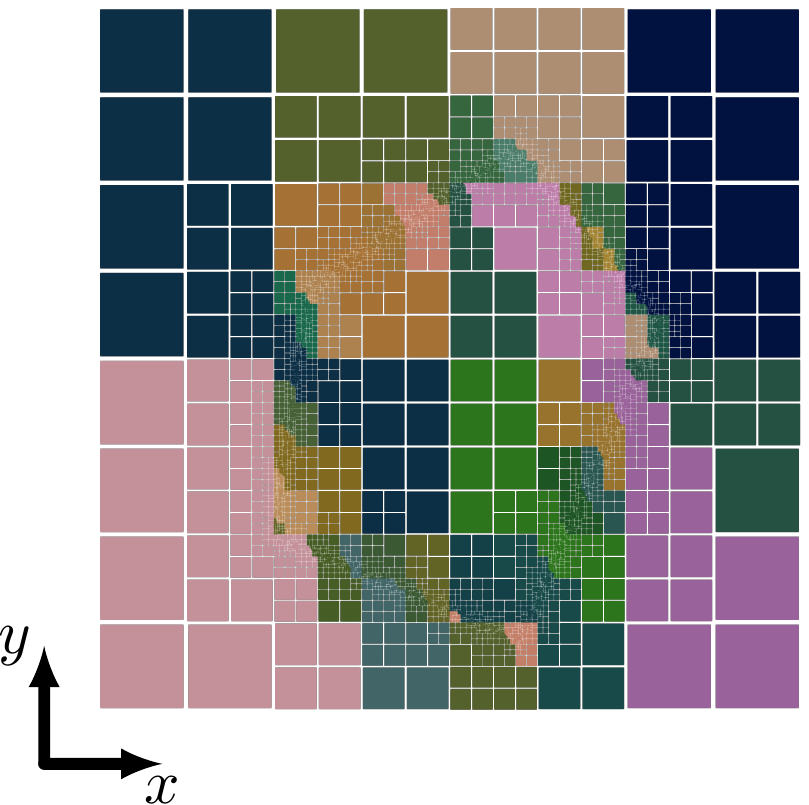}}
	\caption{
		Partitioning of the \pro{} (left) and \cons{} (right) mesh for congruent
		communicators of size $\Ppro = \Pcons = 32$ in the adaptive unit square
		example. The colors encode the partitioning among the processes.
	}
	\label{fig:unit_ranks_adaptive}
	\Description{
		Two paraview-outputs of the adaptively refined, square forests of quadtrees
		from the previous figure showing their partition to 32 ranks.

		The left output shows the partitioning of the producer mesh.
		The right output shows the partitioning of the consumer mesh.
		Each quadrant belongs to exactly one rank.
		Each rank owns no more than two connected areas of quadrants per mesh.
		Most, but not all ranks cover completely different areas in the two meshes.
		The pentagonal area of highly refined quadrants can be roughly made out.
	}
\end{figure}

Both the \cons{} and the \pro{} mesh consist of a single quadtree mapped to the
unit
square.
However, they are mapped in different orientations, as depicted in
\figref{fig:unit_problem_setup}.
Thereby, we make sure that even for congruent communicators the local \cons{}
mesh and \pro{} mesh do not overlap perfectly and thus communication is a part
of our mesh overset tests.
The partitioning of the two meshes is displayed in
\figref{fig:unit_ranks_adaptive}.
Starting with the quadtree being refined uniformly to level 2, we iteratively
refine all quadrants that intersect the boundary of a given pentagon as long as
their level is below 20.
Additionally, we ensure a 2:1 balance condition, by refining adjacent quadrants
when necessary; see
\figref{fig:unit_levels}.

\subsubsection{Numerical Results}\label{sec:unit_square_adaptive_refinement}
We execute the algorithm for 11,432,548 total quadrants per mesh.
We conducted scaling tests on the Marvin MPP cluster for up to 3,072
processes distributed over 32 compute nodes.

\begin{figure}
	\centering
	\subfloat{\includegraphics[width=0.48\textwidth]{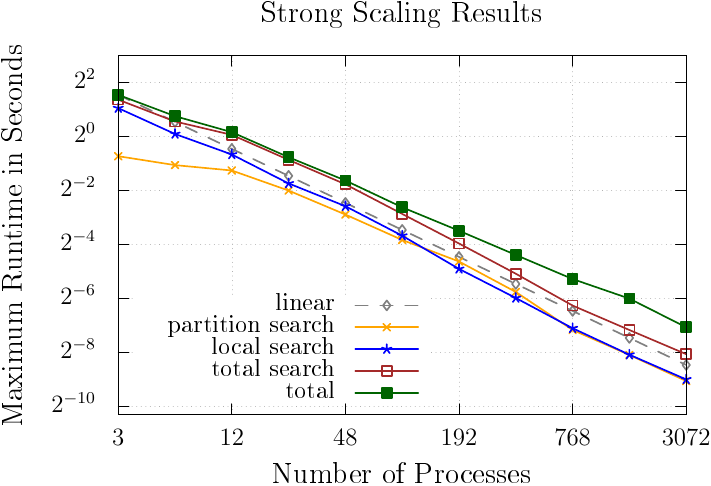}}
	\subfloat{\includegraphics[width=0.48\textwidth]{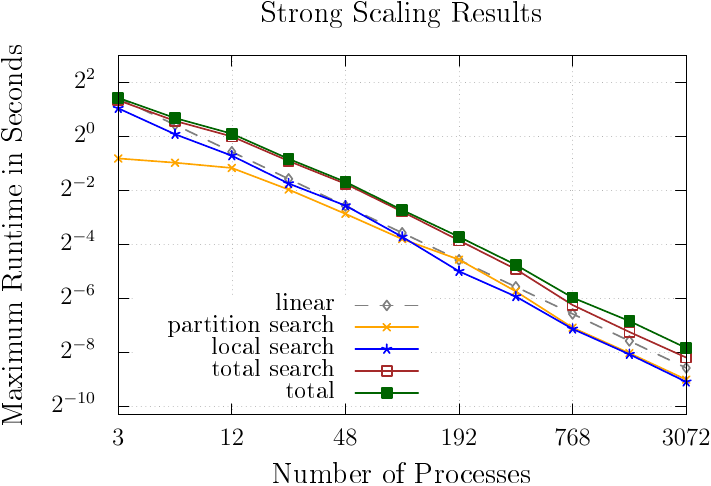}}
	\caption{
		Strong scaling results of the mesh overset for the adaptive unit
		square example (left) and its un-rotated alternative (right) on up to 3,072
		processes.
		The mesh overset is performed on congruent mesh communicators.
		All values represent the maximum time over all processes.
		We display the total run time including waits (green, filled squares),
		the time spent in the partition (orange crosses) and local (blue stars)
		searches (combined: brown squares), as well as linear scaling (gray pluses)
		for reference.
	}
	\label{fig:timings_unit_strong_adaptive}
\Description{
	Two strong scaling plots for the adaptivley refined unit square example in
	the rotated and the unrotated case.

	Both graphics display the maximum run time of the total algorithm as well
	as some of its substeps (partition search, local search, combined searches).
	For comparison theoretical linear scaling is plotted as well.
	The y-axis displays the maximum run time in seconds.
	The x-axis displays the number of processes ranging from 3 to 3072.

	The graph on the left shows the strong scaling for the rotated case.
	From 3 to 12 processes the partition search does not scale leading to
	sub-optimal scaling of the total run time as well.
	From 12 processes to 3072 processes the scalability is slightly worse
	compared to the unrotated case.
	There is a small, but widening gap between total run time and the
	combined time of the searches, which seems to indicate a communication
	overhead.

	The graph on the right shows the strong scaling for the unrotated case.
	From 3 to 12 processes, the partition search does not scale at all.
	This also leads to decreased scalability of the total run time.
	From 12 processes to 3072 processes the scalability is almost optimal.
	The total run time of the mesh overset is almost equal to the combined time
	of the local and partition search.
}
\end{figure}

In order to verify the correctness of the algorithm and its implementation, we
compute the approximation error on the cells.
Since the adaptive refinement is based on the intersection with a pentagon in
physical space both meshes yield the exact same mesh there.
Furthermore, we evaluate an artificial function on the cell centers to
generate `solution' data on the \pro{} side and also use cell centers as queries
on the \cons{} side.
Therefore, we expect the evaluation data we obtain for a query point to exactly
match the solution function when evaluated on its coordinates.
The approximation error in the resulting vector's Euclidean norm indeed
remains below $10^{-12}$ in all test runs.

The results of the strong scaling test are displayed in
\figref{fig:timings_unit_strong_adaptive}.
While for low process counts the increasing complexity of the partition
boundaries still has a significant impact on its scalability, the asymptotic
behavior is almost linear.
The local search scales linearly or even slightly better, due to the recursion
depth decreasing with increasing $\Ppro$.
The combined complexity of both searches scales almost optimally as well.
For low process counts the local search dominates.
With increasing $\Ppro$ the partition search gains relevance.

The overset is subject to a visible communication overhead.
This becomes clear when comparing the test results with the 'un-rotated'
alternative from ~\figref{fig:timings_unit_strong_adaptive} (right), where all
local queries intersect the local \pro{} forest.
Nonetheless, the overall performance of the mesh overset algorithm on this
rather extreme mesh pattern appears reliable.

Since the artificial solution is piecewise constant and the evaluation scheme
relies on a single evaluation point, the time spent on the evaluation part of
the mesh overset is minimized.
Its complexity is proportional to the amount of query points received on the
\pro{} side.
Thus, a more complex evaluation scheme most likely would only improve the
observed scalability.

\subsection{Arc Example}
\label{sec:arcexample}

As a second example we consider the 3D mesh overset of two forests of octrees.
Their domains overlap only partially.
While the \cons{} mesh is shaped like a brick, the \pro{} mesh is shaped like an
arc.
The non-linear mapping of the arc is used to display the viability of the
inverse mapping intersection as described in \secref{sec:intersection_test}.
Furthermore, we will use this example setup to demonstrate how the overset from
\algref{alg:overset}{3} can be used for load balancing of the mesh overset as
well as adaptive refinement at the boundaries of the intersection area.

\subsubsection{Simulation Setup}
\label{sec:arc_simulation_setup}

The \cons{} mesh consists of 6 octrees connected in a $3 \times 2 \times 1$
`brick', with a mapping that scales, rotates and shifts it.
\iffalse
The associated mappings $\phi_0,\ldots,\phi_5$ into physical space are the
composition of two functions.
First, a tree-dependent function shifts the reference domain $[0,1]^3$
of the tree to the corresponding subcube of the $[0,3] \times [0,2] \times
[0,1]$ brick domain.
Second, tree-independent $\phi$ maps into physical space by consecutively
shifting and scaling the brick and rotating it around the $z$-axis as
follows,
\begin{subequations}
\begin{align}
        \phi & \colon [0,3] \times [0,2] \times [0,1] \to \mathbb{R}^3, \\
        \begin{pmatrix}
          u \\
          v \\
          w \\
        \end{pmatrix}
        & \mapsto
        \begin{pmatrix}
          \cos (\frac{\pi}{6}) & - \sin (\frac{\pi}{6}) & 0 \\
          \sin (\frac{\pi}{6}) & \hphantom{-} \cos (\frac{\pi}{6}) & 0 \\
          0 & 0 & 1 \\
        \end{pmatrix}
        \begin{pmatrix}
          \left(u - 1.5 \right) \cdot 0.7 \\
          \left(v - 1 \right) \cdot 0.6   \\
          \left(w - 0.5 \right) \cdot 0.5 \\
        \end{pmatrix}
        +
        \begin{pmatrix}
          0.8 \\
          0.8 \\
          0.5 \\
        \end{pmatrix}.
\end{align}
\end{subequations}
\else
The \pro{} mesh consists of $5$ octrees connected in a $5 \times 1 \times 1$
brick format, which is mapped to form an arc.

We employ an intersection test, which whenever a query point enters a new tree
during the search, applies the tree's inverse mapping and stores the result in
the query point structure.
Thus, it is available for all following intersection tests in the same tree.
For evaluation of the query points we again use an artificial, piecewise
constant solution.

\begin{figure}
	\centering
	\subfloat{\includegraphics[width=0.49\textwidth]{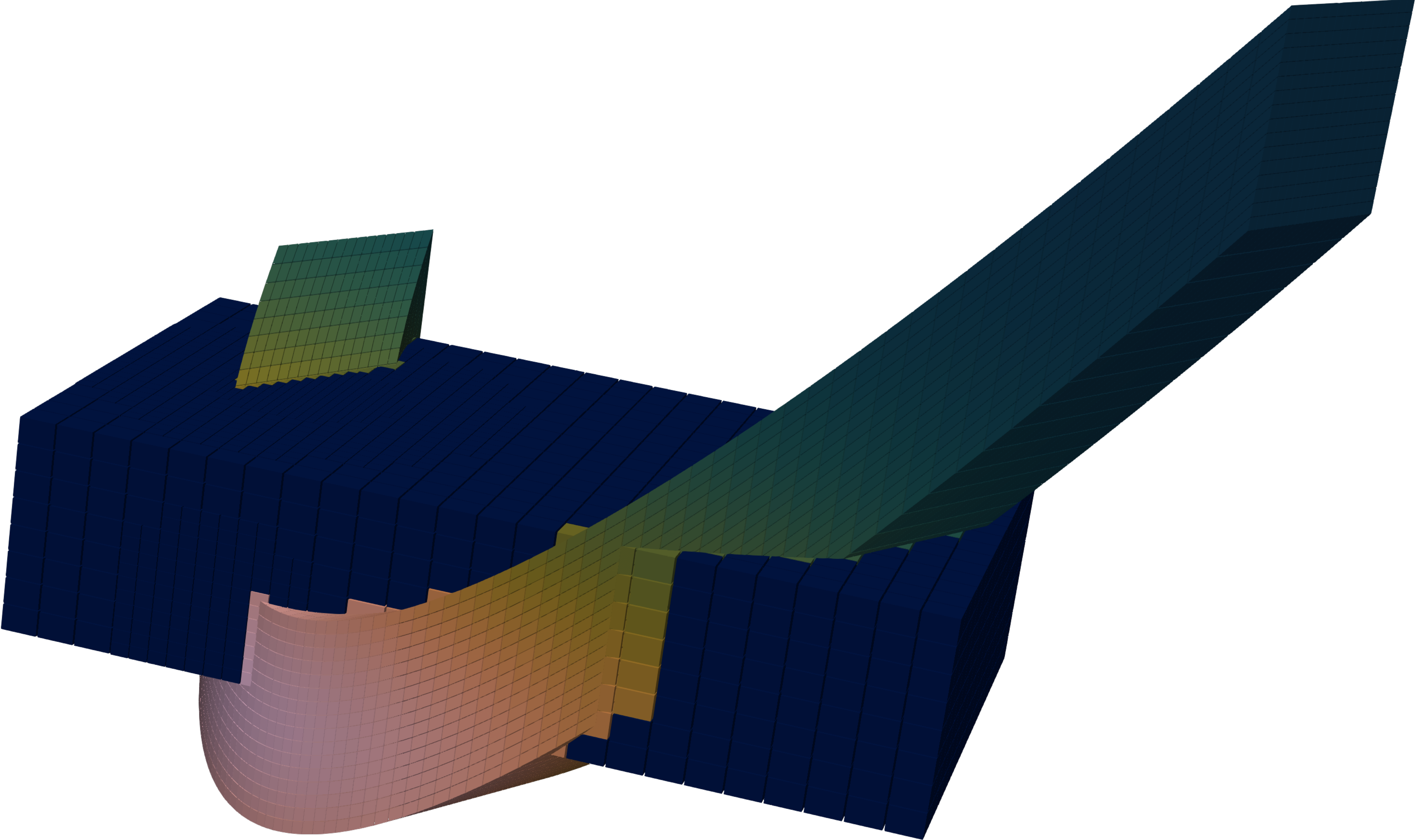}}
	\subfloat{\includegraphics[width=0.49\textwidth]{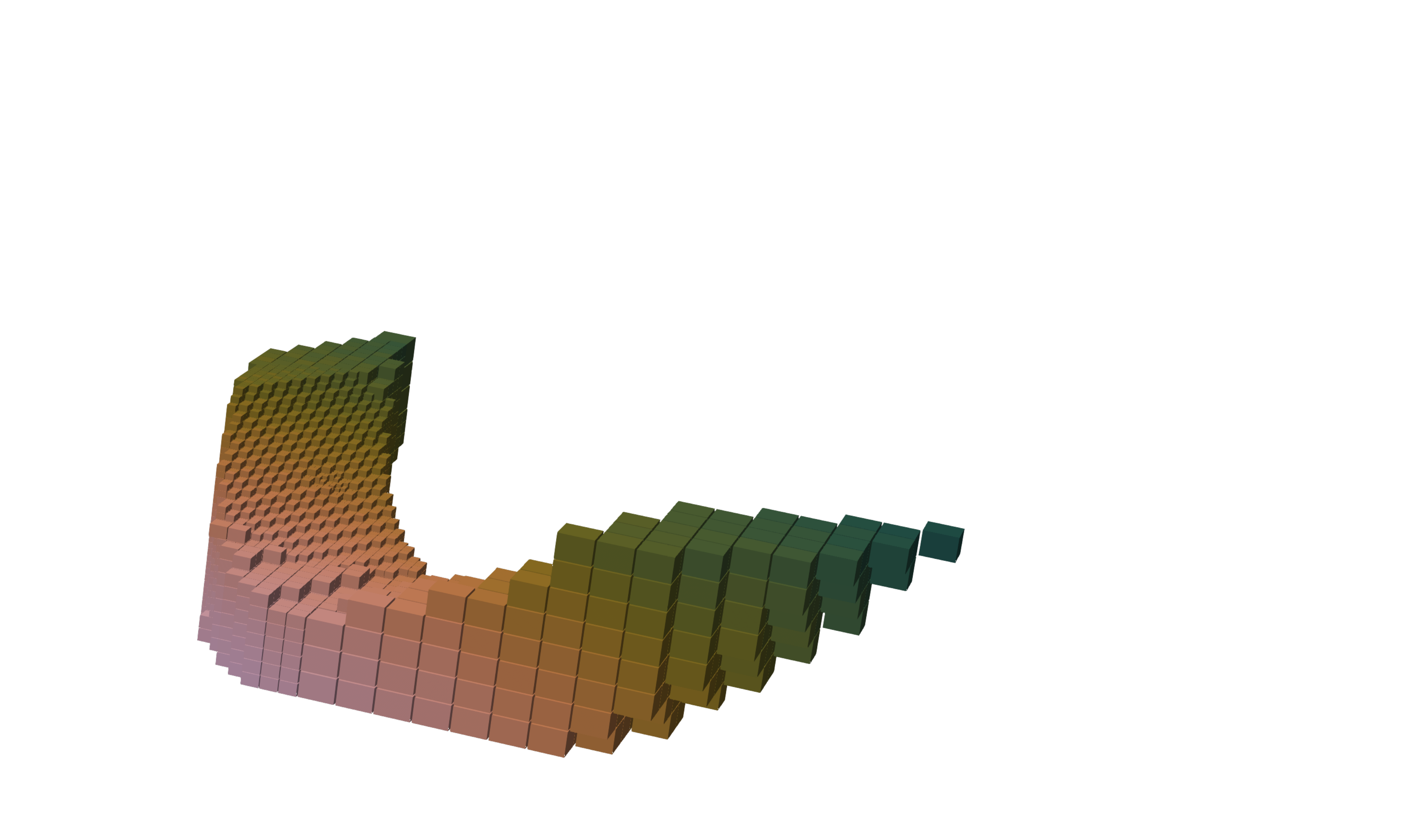}}
	\caption{
		The color-coded evaluation results of the arc example with artificial
		evaluation data.
		The meshes were refined uniformly to level $3$ (\cons{}) and $4$ (\pro{}),
		before being refined adaptively to level $6$.
		We display the \cons{} and \pro{} mesh overlapping each other (left) and the
		\cons{} cells that received evaluation data from the \pro{} (right).
		It can be seen that a higher resolution of the \cons{} mesh results in a
		better approximation of the shape of the arc.
		Equi-distant color scheme taken from~\cite{Crameri23}.
	}
	\label{fig:refinement_solution}
\Description{
	Two visualizations of a 3D mesh overset scenario.

	The left image shows two 3D meshes overlapping each other.
	One mesh is formed like a brick, while the other mesh is formed like an arc.
	The arc mesh takes the role of producer and provides some solution data, which
	is represented by the cells being marked in various different colors.
	The cells of the consumer, brick-like mesh are mostly in the same dark-blue
	color, marking that they did not receive any data from the producer mesh.
	Only the brick cells around the intersection area are colored.
	Their colors matches the color of the clostest producer cells well.

	The right image shows the cells of the brick mesh that received data.
	The resulting shape resembles the expected form of an arc-brick intersection.
	The left half of the adaptively refined brick contains smaller cells than the
	right half.
	Here, the shape of the arc can be made out even clearer.
}
\end{figure}

We refine the \cons{} mesh uniformly to level 7 and the \pro{} mesh uniformly to
level 6 before following up with several iterations of adaptive refinement
into the region of overlap.
As a refinement rule, we compute each octant's distance to an arbitrary fixed
point, which lies inside the intersection area of the meshes, and increase the
level for smaller distances, up to a maximum level of 11.
This results in several concentric spherical shells of different levels and
meshes that are 2:1-balanced by construction.
The \cons{} mesh consists of 2,391,981,241 and the \pro{} mesh of 730,063,983
octants.
In a second, smaller test run we start with uniform levels of 5 and 4 and
refine to a maximum level of 9, resulting in 37,374,497 \cons{} and 11,408,346
\pro{} cells.
An example of the setup can be seen in~\figref{fig:refinement_solution}.

\subsubsection{Numerical Results}\label{sec:arc_numerical_results}
Similar to the previous examples, we perform a strong scaling test with up
to 12,288 processes on 128 nodes of the Marvin MPP cluster.

Again, we start by verifying the correctness of the algorithm and its
implementation.
Unlike for the unit square example we can no longer assume that we will end
up with exact matches between the interpolated overset results and the exact
solution data.
Instead, we compute a relative error by dividing the norm of the vector
of solutions at every query point through the norm of the vector of errors,
%
\begin{table}
	\begin{center}
		\begin{tabular}{cccccc}
			\toprule
			maxlevel $m$ & 7 & 8 & 9 & 10 & 11\\
			\midrule
			relative $\ell_\infty$ error & 0.1251 & 0.0612 & 0.0321 & 0.0168 & 0.0086\\
			ratio $m-1$ to $m$ & -- & 2.044 & 1.907 & 1.911 & 1.953 \\
			\midrule
			relative $\ell_2$ error & 0.00720 & 0.00341 & 0.00168 & 0.00083 & 0.00041\\
			ratio $m-1$ to $m$ & -- & 2.111 & 2.030 & 2.024 & 2.024 \\
			\bottomrule
		\end{tabular}%
	\end{center}%
	\caption{Relative error of the mesh overset based interpolation in the
		$\ell_\infty$ and the $\ell_2$ norm for different maximum level $m$.
		The decrease of the error from one level to another is shown as well.}%
	\label{tab:relative_error}
\end{table}%
%
which meets our expectations as depicted in
Table~\ref{tab:relative_error}.

\begin{figure}
	\centering
	\includegraphics[width=.55\textwidth]{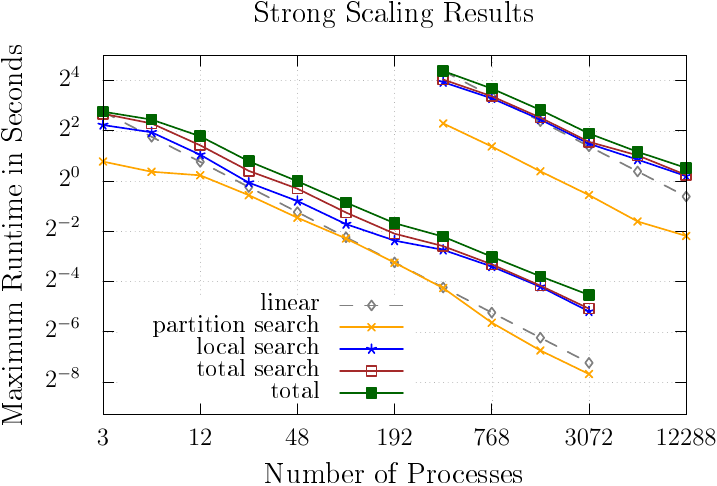}
	\hspace{0.8cm}
	\caption{
		Strong scaling results of the mesh overset for the arc example without load
		balancing on up to 12,288 processes.
		The mesh overset is performed on congruent mesh communicators.
		We display the maximum run time of the complete mesh overset and some of
		its substeps.
		The graphs show total run time (green, filled squares), the time spent in
		the partition (orange crosses) and local (blue stars) searches (combined:
		brown squares), as well as linear scaling (gray pluses).
	}
	\label{fig:timings_arc_strong}
\Description{
	Strong scaling plot for the adaptively refined arc example showing
	maximum run times.

	The graphic displays the run time of the total algorithm as well as some of
	its substeps (partition search, local search, combined searches).
	For comparison theoretical linear scaling is plotted as well.
	The y-axis displays the run time in seconds.
	The x-axis displays the number of processes ranging from 3 to 12288.
	In the range from 3 to 3072 processes a small test was performed.
	In the range from 384 to 12288 a second test was performed for a
	larger problem size.

	The graph shows the maximum run time of the strong scaling test.
	Similar to the scaling results shown in previous sections, the partition
	search does not scale well from 3 to 12 processes, but scales almost linearly
	from 12 to 3072 processes.
	The maximum run time of the local search does not scale well in the whole
	range from 3 to 3072 processes.
	This also leads to a mediocre scalability of the algorithm as a whole.
	For the second, larger run, the overall scalability is better, but not yet
	optimal.
}
\end{figure}

The results of the strong scaling test are displayed in
\figref{fig:timings_arc_strong}.
Similar to the 2D unit square example from
\secref{sec:unit_square_adaptive_refinement} the partition search does not scale
well going from 3 to 12 processes, but recovers for higher process counts.

The local search however does not scale well anywhere going from 3 to 12,228
processes.
This can be explained with the increasing load imbalance in this first version
of our problem setup.
As can be seen in \figref{fig:refinement_solution}, large parts of both the
\cons{} and the \pro{} mesh are not part of their common intersection area.
Processes only owning octants outside the overlap region will not contribute
to the mesh overset at all.
This effect can also be measured by looking at the maximum number of query
points any \pro{} process receives during the overset and how it scales (see
\figref{fig:timings_arc_queries_strong}).
While for three processes we have a maximum of 9,609,023 query points with a
standard deviation of $10.1\%$, for 3,072 processes we have a maximum of
115,512 queries with a standard deviation of $103\%$.
So, the maximum number of points to be searched on a single process reduces only
by a factor of 83, while the number of processes increases by a factor of 1,024.
Thus, the scalability of the mesh overset is limited by the problem's inherent
load imbalances.
Luckily, we can tackle load imbalances by a weighted repartitioning
approach, as described in the next section.

\subsection{Load Balancing}\label{sec:load_balancing}
The arc example as described in \secref{sec:arc_simulation_setup} is subject to
significant, problem-inherent load imbalance.
Large parts of both the \cons{} and the \pro{} mesh do not contribute to the
intersection area at all, while others contain a mass of high-level
octants.

Commonly, parallel mesh overset methods mitigate load imbalance by changing
the distribution of grid sub-blocks, e.g.\ based on
bin-packing~\cite{HenshawSchwendeman08} or splitting of the grid sub-blocks
into smaller parts~\cite{RogetSitaraman14}.
This cannot be done in our mesh overset setup, since the space-filling curve
partitioning of the forest of octrees is native to the entire logic.
Shifting the partition boundaries in the linear order of octants is the
only option, yet a powerful one:
%
To work towards optimal load balance, we perform a weighted
repartitioning~\cite{BursteddeWilcoxGhattas11} of both the \cons{} and the
\pro{}
forest of octrees by assigning a weight to every leaf octant representing its
workload.

We detail our approach as follows.
The workload associated with the searches depends strongly on the amount of
octant-point intersection callbacks they have to perform.
The partition search is executed on the \cons{} side of the overset algorithm.
The partition boundaries and thereby the amount of octants to traverse during
the search is the same on all processes and cannot be reduced conveniently.
However, the amount of query points entering the search depends on the local
\cons{} octants and is a target well-suited for load balancing.

During the partition search query points that do not intersect the \pro{} domain
are part of one intersection test for each of the $5$ \pro{} tree roots, before
being rejected.
Query points that intersect the \pro{} domain are searched until they can be
assigned to a unique process, resulting in $8$ (or $4$ in 2D) intersection
tests per search level.
To account for this, we add $1$ to a leaf octant's weight for every query point
outside the \pro{} domain and a heuristically chosen weight of $7$ for each
query
point inside the \pro{} domain.

The local search is executed on the \pro{} side of the overset algorithm.
The amount of points entering the search depends on the local \pro{} octants.
Every query point contained in a given \pro{} local leaf octant has to pass
through all previous levels of the local forest, resulting in $8$ (or $4$ in
2D) intersection tests per search level.
We add a heuristically chosen weight of $20$ for every query point contained in
an octant to its base weight of $1$.

To compute the \cons{} and \pro{} octant weights, we strive to determine which
query points intersect the \pro{} domain and how many query points intersect
any given \pro{} octant.
This can be achieved by applying the overset algorithm.
We use the same
\fxn{Intersection\_Test} as for a normal mesh overset routine.
For evaluation, we use a new callback \fxn{Evaluate\_Load\_Balancing}, which
marks the query point to be inside the intersection area and increments the
number of queries found in the \pro{} cell.

To test the effect of our heuristic load balancing, we apply it to the test
case from \secref{sec:arc_numerical_results} and compare the results.
The strong scaling of the load balanced run can be seen in
\figref{fig:timings_arc_strong_load_balance}.
\begin{figure}
	\centering
	\includegraphics[width=.55\textwidth]{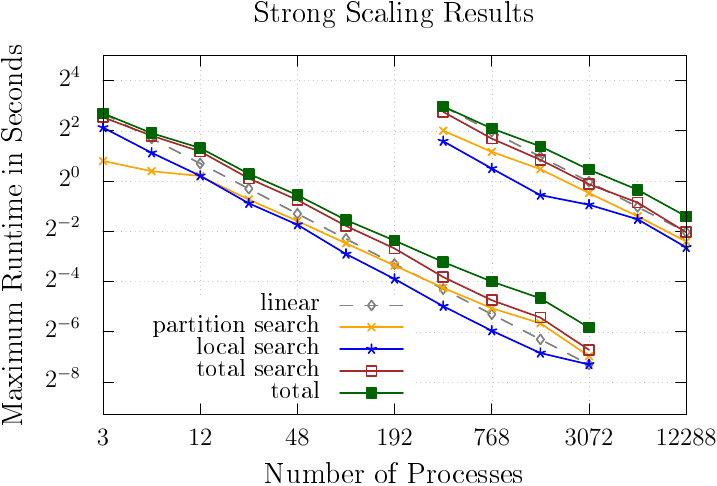}
	\hspace{0.8cm}
%
	\caption{
		Strong scaling results of the load-balanced mesh overset for the arc
		example on up to 12,288 processes.
		The mesh overset is performed on congruent mesh communicators.
		We display the maximum run time of the complete mesh overset and some of
		its substeps.
		The graphs show total run time (green, filled squares), the time spent in
		the partition (orange crosses) and local (blue stars) searches (combined:
		brown squares), as well as linear scaling (gray pluses).
		The overall run time exhibits significantly improved scalability compared
		to the unbalanced results from \figref{fig:timings_arc_strong}.
		For example, the total run time for 12,288 processes dropped from 1,44s to
		0,38s.
	}
	\label{fig:timings_arc_strong_load_balance}
\Description{
	Strong scaling plot for the adaptively refined arc example showing
	maximum run times.

	The graphic displays the run time of the total algorithm as well as some of
	its substeps (partition search, local search, combined searches).
	For comparison theoretical linear scaling is plotted as well.
	The y-axis displays the run time in seconds.
	The x-axis displays the number of processes ranging from 3 to 12288.
	In the range from 3 to 3072 processes a small test was performed.
	In the range from 384 to 12288 processes a second test was performed for a
	larger problem size.

	The graph shows the maximum run time of the strong scaling test.
	Similar to the scaling results shown in previous sections, the partition
	search does not scale well from 3 to 12 processes, but scales almost linearly
	from 12 to 3072 processes.
	The maximum run time of the local search does scale well in the whole
	range from 3 to 3072 processes, except for a light bump for 3072 processes.
	The overall mesh overset scales decently as well.
	For small process counts the total run time is almost identical to the total
	time spent in searches.
	For larger process counts there is a visible gap most likely caused by
	communication.
	For the second, larger run, the overall scalability is similar.
}
\end{figure}
\begin{figure}
	\centering
	\subfloat{\includegraphics[width=0.45\textwidth]{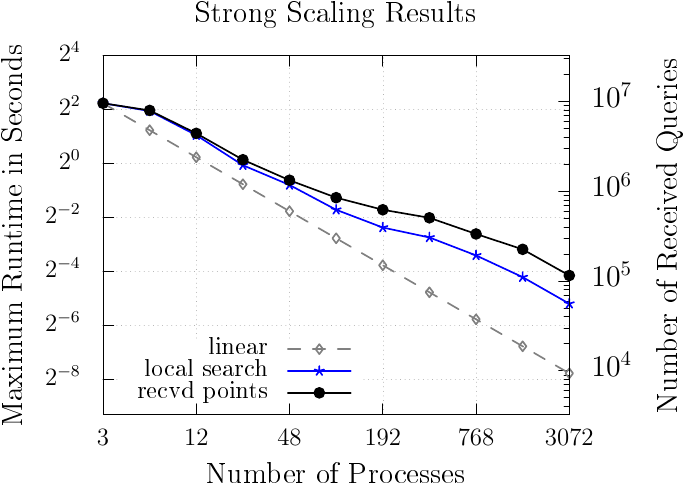}}
	\hspace{1cm}
	\subfloat{\includegraphics[width=0.45\textwidth]{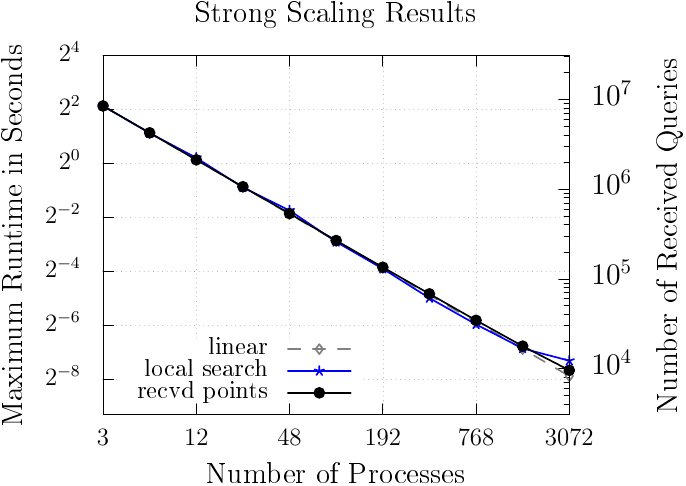}}
	\caption{
		Comparison of the local search run times of the mesh overset in the
		non load-balanced (left) and the load-balanced (right) arc example on up to
		3,072 processes.
		The mesh overset is performed on congruent mesh communicators.
		We display the maximum run time of the local search (blue stars) as well as
		the maximum number of query points received on any \pro{} process (black
		circles) and theoretical linear scaling (gray pluses).
		The significantly improved scalability of the local search underlines the
		need for load balancing.
	}
	\label{fig:timings_arc_queries_strong}
\Description{
	Two strong scaling plots for the arc example showing maximum run time of
	the local search for the non load-balanced and the load-balanced overset.

	Both graphics display the maximum run time of the local search as well as the
	maximum number of points received on every process.
	For comparison theoretical linear scaling is plotted as well.
	The y-axis displays the run time in seconds on the left and number of points
	received on the right.
	The x-axis displays the number of processes ranging from 3 to 3072.
	The y-axes are scaled such that the number of received queries and the total
	run time start at the same point for the minimum of 3 processes, which allows
	to compare their scalability.

	The graph on the left shows the results for the non load-balanced mesh
	overset.
	Here, the local search scales poorly almost everywhere from 3 to 3072
	processes.
	The maximum number of received queries describes a very similar curve, but
	scales worse for higher process counts.

	The graph on the right shows the results for the load-balanced mesh overset.
	Here, both the local search and the number of received queries scale linearly
	everywhere from 3 to 3072 processes, except for a small bump for 3072
	processes.
}
\end{figure}%
The scalability of the mesh overset improves visibly, which also comes with a
drop in total run times.
For example, the time to compute the overset for $P=384$ in the larger problem
setup reduces from 20,9s to 7,9s.
In particular, the local search---which was the main reason for the
far-from-optimal results in the unbalanced overset---now scales almost
linearly.
As can be seen in \figref{fig:timings_arc_queries_strong}, our repartitioning
approach yields the desired results and leads to linear scalability of the
maximum amount of queries received on any process.

The improved scalability of the load-balanced overset justifies performing a
weak scaling test.
Whenever we increase the amount of processes by a factor of eight, we increase
the initial uniform level as well as the maximum refinement level by one.
The results can be seen in \figref{fig:timings_arc_weak_load_balance}.
For comparison, the graph contains a heuristic approximation of the workload
of searching the the \cons{} cells in the \pro{} forest.

\begin{figure}
	\centering
	\subfloat{\includegraphics[width=0.48\textwidth]{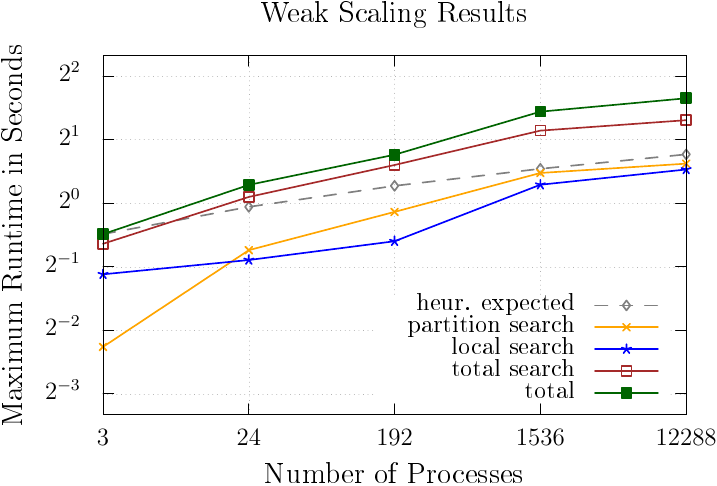}}
	\hfill
	\subfloat{\includegraphics[width=0.48\textwidth]{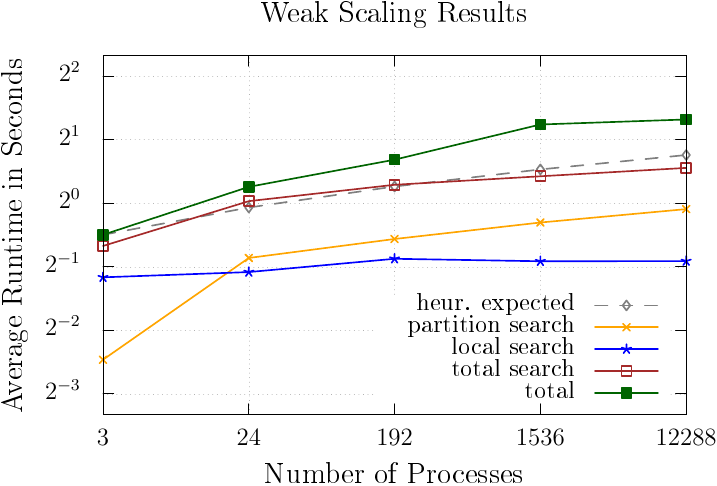}}
	\caption{
		Weak scaling results of the load-balanced mesh overset for the arc
		example on up to 12,288 processes.
		The mesh overset is performed on congruent mesh communicators.
		We display the maximum run time of the complete mesh overset and some of
		its substeps in (left) and the average run time in (right).
		The graphs show total run time (green, filled squares), the time spent in
		the partition (orange crosses) and local (blue stars) searches (combined:
		brown squares), as well as heuristically expected scaling (gray pluses).
	}
	\label{fig:timings_arc_weak_load_balance}
\Description{
	Two weak scaling plots for the load-balanced arc example showing maximum and
	average run times.

	Both graphics display the run time of the total algorithm as well as some of
	its substeps (partition search, local search, combined searches).
	For comparison the heuristically expected scaling is plotted, which expects a
	slight increase in run time for larger process counts.
	The y-axis displays the run time in seconds.
	The x-axis displays the number of processes ranging from 3 to 12288.

	The graph on the left shows the maximum run time of the weak scaling test.
	The total run time scales decently, but there is a visible gap to the expected
	scaling which widens with increasing process counts.
	The total time spent in search routines takes a very similar path to the total
	run time.
	The local search scales almost linearly for the first three and the last
	measurement, but slows down strongly going from 192 to 1536 processes.
	The partition search scales poorly at first, but improves steadily.

	The graph on the right shows the average run time of the weak scaling test.
	The overall run time still scales decently, but with visible room for
	improvement.
	The total run time scales almost exactly as it would be heuristically
	expected.
	The average run time of the local search is completely straight, while the
	partition search steadily slows down.
}
\end{figure}
The weak scalability
of the
partition search 
remains the least predictable, which is to be
expected, as the partition boundaries get more complex with every step.
The local search scales linearly on average.


\subsection{Adaptive Refinement Around the Intersection Area}\label{sec:adaptive_refinement}

Proper transmission of the solution from one mesh to the other requires sufficient resolution of the receiving mesh --
in particular in the intersection area or on its boundary~\cite{KirbyBrazellYangEtAl19}.
This can be ensured by adaptively refining the mesh around the intersection area.
The overset algorithm presented in this paper can also be used for such
refinement, which we demonstrate by refining both the \pro{} and the \cons{}
forest
of octrees on the boundary of the intersection area.

We start with marking all boundary octants of the meshes using a flag $\oc{o.isBoundary}$.
Our approach is to set a refinement flag $\oc{o.refine}$ for all \pro{} and
\cons{} leaf octants $\oc o$.
We want to set it to \algtrue{} if and only if
\begin{itemize}
	\item $\oc o$ intersects an octant from the other mesh and
	\item at least one of the two octants is a boundary octant in its mesh.
\end{itemize}
We can not check this criterion precisely without implementing an intersection test for a pair of mapped octants from both meshes, which might cause high theoretical and/or computational effort.
Instead, we choose a heuristic approach:
We define a set of query points for every consumer cell according to a $3 \times 3 \times 3$-tensor product.
We query the octant's corners, and the centers of all its faces, edges and volume.
Every query point \qpoint is supplied with its coordinates and the flag $\qpoint \oc{.isBoundary}$ inherited from $\oc{o.isBoundary}$.
\begin{algo}
	\hypertarget{alg:evaluate_refinement}{}
	\includegraphics{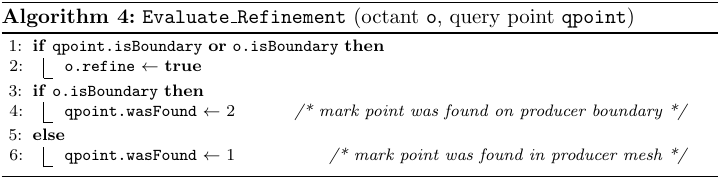}
\end{algo}

This setup allows us to use \fxn{Overset} for the refinement by choosing
suitable callbacks.
We use the same \fxn{Intersection\_Test} callback as for a normal mesh overset routine.
For evaluation, we use a new callback \fxn{Evaluate\_Refinement} (see
\algref{alg:evaluate_refinement}{4}), which marks both the query point and the
octant according to our criterion.
After the overset is completed, we check the updated query points for every
\cons{} octant and mark the octant for refinement accordingly.
The whole procedure can be seen in \algref{alg:set_refinement_flags}{5}.
For simplicity, we consider only congruent communicators, where both the \pro{}
forest $f_p$ and the \cons{} forest $g_p$ are available locally.
\begin{algo}
	\hypertarget{alg:set_refinement_flags}{}
	\includegraphics{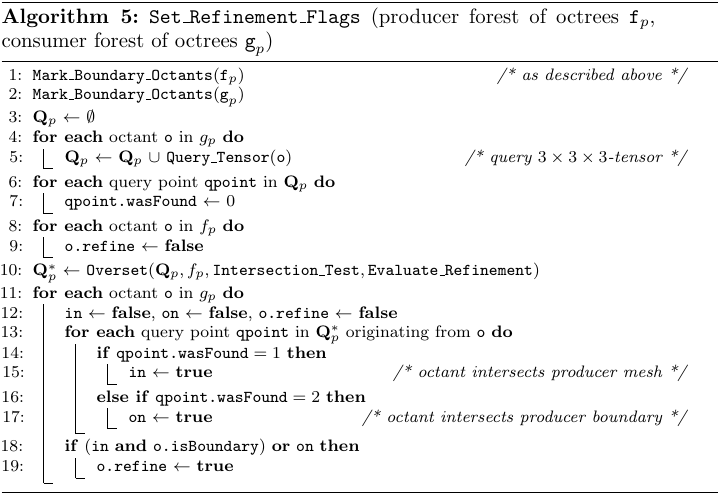}
\end{algo}

We test our approach, by adaptively refining the \cons{} brick and \pro{} arc
from level $2$ to a maximum level $m = 12$ on the boundary of the
intersection area.
To test our approach, we start with a \cons{} brick and a \pro{} arc both
refined
uniformly to level
$2$ and use the procedure described above to identify and refine all octants
below level $12$ on the boundary of the intersection area by one level.
We repeat this process iteratively, until no octants are refined anymore.

\begin{figure}
	\centering
	\includegraphics[width=\textwidth]{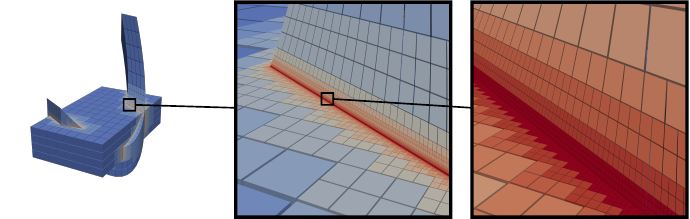}
	\caption{
		Color-coded levels of the adaptively refined arc example.
		From left to right, the images subsequently zoom in onto a part of
		the boundary of the intersection area, as indicated by the black boxes.
		The colors represent the different levels going from $2$ (blue) to $12$
		(red).
	}
	\label{fig:adaptive_refinement}
\Description{
Three images displaying the resulting mesh levels of the adaptive mesh
refinement based on the overset.

There are three consecutive images of the brick and the arc mesh
from previous examples.
Their cells are colored according to the mesh level.
The three images form a sequence that zooms in on one edge of the intersection
area of arc and brick.
It can be seen, that both meshes are refined up to a very fine level at this
edge.
Around the edge the cells quickly get much coarser while maintaining a 2:1
balance.
}
\end{figure}

The refinement procedure gives a good approximation of the intersection area, as can be seen in \figref{fig:adaptive_refinement}.
For further verification we alter the maximum refinement level $m$ and compare the results.
The boundary of the intersection area has a fractal dimension of $2$.
Thus, we would expect the number of maximum level octants to quadruple,
when incrementing the maximum level by one.
The actual increase in maximum level octants observed during the test runs
can be seen in Table~\ref{tab:maxlevel_quads}.
The observed ratios match the expectations well.
With increasing maximum level, the surface of the intersection area can be
represented with higher accuracy in both meshes, resulting in a ratio even
closer to $4$.

\begin{table}
\begin{center}
\begin{tabular}{cccccc}
		\toprule
		$m$ & 5 & 6 & 7 & 8 & 9 \\
		\midrule
		\# $\lbrace$ cons. octants of level $m$ $\rbrace$ & 28,888 & 124,496 & 515,832 & 2,099,096 & 8,507,520 \\
		ratio $m$ to $m - 1$ & -- & 4.31 & 4.14 & 4.07 & 4.05 \\
		\midrule
		\# $\lbrace$ prod. octants of level $m$ $\rbrace$& 36,696 & 157,496 & 646,336 & 2,589,006 & 10,407,664 \\
		ratio $m$ to $m - 1$ & -- & 4,29 & 4.10 & 4.01 & 4.02 \\
		\bottomrule
\end{tabular}%
\end{center}%
\caption{Number of \cons{} and \pro{} octants of maximum level $m$ for adaptive
         refinement around the intersection area in the arc example.
         The results are displayed for $m$ ranging from 5 to 9.
         The increase of maximum level octants between the run of level $m$
         and the run of level $m-1$ is shown as well.}%
\label{tab:maxlevel_quads}
\end{table}%

\subsection{Geophysics Simulations}
Next to the artificial examples discussed in the previous sections, the
\fxn{Overset} algorithm has already been used for real-world applications like
mesh coupling in acoustic gravity waves (AGWs) simulations~\cite{SnivelyCalhounAitonEtAl24,SnivelyCalhounAitonEtAl24a,SnivelyCalhounAitonEtAl23, ZettergrenCalhounSnivelyEtAl25, SnivelyCalhounAitonEtAl25, SnivelyCalhounAitonEtAl25a}.
The resulting simulation model was scaled to 16,384 cores.

The behavior of the AGWs near their origin on earth's surface can be simulated
by the Model for Acoustic-Gravity wave Interactions and Coupling
(MAGIC)~\cite{Snively13}.
The behavior in the ionosphere, where AGWs are ultimately observed, can
subsequently be modeled using the Geospace Environment Model for Ion-Neutral
Interactions (GEMINI)~\cite{ZettergrenSnively15}.
Both models rely on meshes from the
\texttt{ForestClaw}~\cite{CalhounBurstedde17} library, which itself inherits
its adaptive mesh refinement capabilities from \pforest{}.
In practice, the MAGIC and GEMINI meshes were be coupled efficiently using an
implementation of this algorithm in \texttt{ForestClaw}.
GEMINI assumes the role of \cons{} and creates a set of query points for each
cell in its curved domain.
The queries are supplied with interpolation data by MAGIC, which is the \pro{}
in
this scenario.

\section{Conclusion}
In this work, we introduce a new algorithm for the mesh overset of a parallel
distributed forest of octrees with another mesh of unrelated partition.
The second mesh is defined only by parallel distributed sets of query
points.
Thus, it may be of almost arbitrary structure.
The forest of octrees consists of one or multiple trees, each supplied with its
own smooth mapping.
Thereby, it can represent a broad range of geometric domains.

We leverage a communication-free, exact partition search of the forest of
octrees to match every query point with the
process containing it.
Based on the results of this search, the query points are sent to and returned
from the respective processes.
On the leaf level the query points are supplied with the queried data.
While the algorithm was developed for a one-directional exchange, a
two-directional overset between two forest-of-octree meshes can also be realized
by applying the presented method once in each direction,
with synergies in message aggregation.


One key aspect of our design is its modularity; since
the query points are treated as opaque structures that are passed to
problem-dependent callback functions for intersection tests and evaluation.
Thus, a query `point' may also be a query `cell' or query `object'.
This abstraction allows us to extend the proposed algorithm to a variety of
overset-related problems covered in this paper, i.e.\ adaptive refinement
around the intersection area and load balancing via weighted repartitioning,
and hints at applications in computational geometry.

We find that the algorithm is quick and scales well to 12,288 cores on the
Marvin MPP cluster, in particular when combined with the SFC-based
repartitioning of the cells for load balancing.
%
%
This makes it an excellent candidate for addressing more involved,
multi-mesh overset setups.
%
%
For example, instead of using a forest of octrees as one of the
participating meshes, it might be used as a distributed-memory background
structure to organize the overset of multiple (unstructured) meshes.

\section*{Acknowledgments}
We would like to thank Donna Calhoun and Scott Aiton for their support in
incorporating this
mesh overset algorithm into the \texttt{ForestClaw} software library, and
Jonathan Snively as well as Matthew Zettergren for applying and thereby
testing the resulting code in a real-world geophysics simulation based on
the \texttt{MAGIC} and \texttt{GEMINI} codes.
We gratefully appreciate their continuous and insightful feedback on our
algorithm and its interface.
Furthermore, we would like to thank the support team of the Marvin cluster,
above all Dirk Barbi, Jan Steiner and Florian Boecker, for the optimization
of the Marvin runtime environment, which made the scaling tests possible.

\bibliographystyle{ACM-Reference-Format}
\bibliography{../group,../ccgo_new}


\begin{thebibliography}{38}


\ifx \showCODEN    \undefined \def \showCODEN     #1{\unskip}     \fi
\ifx \showDOI      \undefined \def \showDOI       #1{#1}\fi
\ifx \showISBNx    \undefined \def \showISBNx     #1{\unskip}     \fi
\ifx \showISBNxiii \undefined \def \showISBNxiii  #1{\unskip}     \fi
\ifx \showISSN     \undefined \def \showISSN      #1{\unskip}     \fi
\ifx \showLCCN     \undefined \def \showLCCN      #1{\unskip}     \fi
\ifx \shownote     \undefined \def \shownote      #1{#1}          \fi
\ifx \showarticletitle \undefined \def \showarticletitle #1{#1}   \fi
\ifx \showURL      \undefined \def \showURL       {\relax}        \fi
\providecommand\bibfield[2]{#2}
\providecommand\bibinfo[2]{#2}
\providecommand\natexlab[1]{#1}
\providecommand\showeprint[2][]{arXiv:#2}

\bibitem[Burstedde(2010)]%
        {Burstedde25}
\bibfield{author}{\bibinfo{person}{Carsten Burstedde}.}
  \bibinfo{year}{2010}\natexlab{}.
\newblock \bibinfo{title}{{\texttt{\upshape p4est}}: Parallel {AMR} on Forests
  of Octrees}.
\newblock
\newblock
\newblock
\shownote{\url{https://www.p4est.org/} (last accessed December 28th, 2025)}.


\bibitem[Burstedde(2020)]%
        {Burstedde20d}
\bibfield{author}{\bibinfo{person}{Carsten Burstedde}.}
  \bibinfo{year}{2020}\natexlab{}.
\newblock \showarticletitle{Parallel tree algorithms for {AMR} and non-standard
  data access}.
\newblock \bibinfo{journal}{\emph{ACM Trans. Math. Software}}
  \bibinfo{volume}{46}, \bibinfo{number}{32} (\bibinfo{date}{November}
  \bibinfo{year}{2020}), \bibinfo{pages}{1--31}.
\newblock
Issue 4.
\urldef\tempurl%
\url{https://doi.org/10.1145/3401990}
\showDOI{\tempurl}


\bibitem[Burstedde et~al\mbox{.}(2013)]%
        {BursteddeStadlerAlisicEtAl13}
\bibfield{author}{\bibinfo{person}{Carsten Burstedde}, \bibinfo{person}{Georg
  Stadler}, \bibinfo{person}{Laura Alisic}, \bibinfo{person}{Lucas~C. Wilcox},
  \bibinfo{person}{Eh Tan}, \bibinfo{person}{Michael Gurnis}, {and}
  \bibinfo{person}{Omar Ghattas}.} \bibinfo{year}{2013}\natexlab{}.
\newblock \showarticletitle{Large-scale adaptive mantle convection simulation}.
\newblock \bibinfo{journal}{\emph{Geophysical Journal International}}
  \bibinfo{volume}{192}, \bibinfo{number}{3} (\bibinfo{year}{2013}),
  \bibinfo{pages}{889--906}.
\newblock
\urldef\tempurl%
\url{https://doi.org/10.1093/gji/ggs070}
\showDOI{\tempurl}


\bibitem[Burstedde et~al\mbox{.}(2011)]%
        {BursteddeWilcoxGhattas11}
\bibfield{author}{\bibinfo{person}{Carsten Burstedde},
  \bibinfo{person}{Lucas~C. Wilcox}, {and} \bibinfo{person}{Omar Ghattas}.}
  \bibinfo{year}{2011}\natexlab{}.
\newblock \showarticletitle{{\texttt{\upshape p4est}}: Scalable Algorithms for
  Parallel Adaptive Mesh Refinement on Forests of Octrees}.
\newblock \bibinfo{journal}{\emph{SIAM Journal on Scientific Computing}}
  \bibinfo{volume}{33}, \bibinfo{number}{3} (\bibinfo{year}{2011}),
  \bibinfo{pages}{1103--1133}.
\newblock
\urldef\tempurl%
\url{https://doi.org/10.1137/100791634}
\showDOI{\tempurl}


\bibitem[Burstedde et~al\mbox{.}(2010)]%
        {BursteddeWilcoxIsaac25}
\bibfield{author}{\bibinfo{person}{Carsten Burstedde},
  \bibinfo{person}{Lucas~C. Wilcox}, {and} \bibinfo{person}{Tobin Issac}.}
  \bibinfo{year}{2010}\natexlab{}.
\newblock \bibinfo{title}{The "p4est" forest-of-octrees library}.
\newblock
\newblock
\newblock
\shownote{\url{https://github.com/cburstedde/p4est} (last accessed October
  24th, 2025)}.


\bibitem[Calhoun and Burstedde(2017)]%
        {CalhounBurstedde17}
\bibfield{author}{\bibinfo{person}{Donna Calhoun} {and}
  \bibinfo{person}{Carsten Burstedde}.} \bibinfo{year}{2017}\natexlab{}.
\newblock \bibinfo{title}{{ForestClaw}: A parallel finite volume solver for
  adaptively refined {C}artesian multiblock domains}.  (\bibinfo{year}{2017}).
\newblock
\showeprint[arxiv]{1703.0317}~[cs.MS]
\newblock
\shownote{Manuscript}.


\bibitem[Crameri(2023)]%
        {Crameri23}
\bibfield{author}{\bibinfo{person}{Fabio Crameri}.}
  \bibinfo{year}{2023}\natexlab{}.
\newblock \bibinfo{booktitle}{\emph{Scientific colour maps}}.
\newblock Fabio Crameri.
\newblock
\urldef\tempurl%
\url{https://doi.org/10.5281/zenodo.8409685}
\showDOI{\tempurl}


\bibitem[Gagnon(2022)]%
        {Gagnon22}
\bibfield{author}{\bibinfo{person}{Micha{\"e}l Gagnon}.}
  \bibinfo{year}{2022}\natexlab{}.
\newblock \emph{\bibinfo{title}{Development of a Parallel 3D Solid-Solid
  Overset {RANS} Method}}.
\newblock \bibinfo{thesistype}{Master's\ thesis}.
  \bibinfo{school}{Polytechnique Montr{\'e}al}.
\newblock
\urldef\tempurl%
\url{https://publications.polymtl.ca/10351/}
\showURL{%
\tempurl}


\bibitem[Hedayat et~al\mbox{.}(2022)]%
        {HedayatAkbarzadehBorazjani22}
\bibfield{author}{\bibinfo{person}{Mohammadali Hedayat},
  \bibinfo{person}{Amir~M. Akbarzadeh}, {and} \bibinfo{person}{Iman
  Borazjani}.} \bibinfo{year}{2022}\natexlab{}.
\newblock \showarticletitle{A Parallel Dynamic Overset Grid Framework for
  Immersed Boundary Methods}.
\newblock \bibinfo{journal}{\emph{Computers \& Fluids}}  \bibinfo{volume}{239}
  (\bibinfo{year}{2022}), \bibinfo{pages}{105378}.
\newblock
\showISSN{0045-7930}
\urldef\tempurl%
\url{https://doi.org/10.1016/j.compfluid.2022.105378}
\showDOI{\tempurl}


\bibitem[Henshaw and Schwendeman(2008)]%
        {HenshawSchwendeman08}
\bibfield{author}{\bibinfo{person}{William~D. Henshaw} {and}
  \bibinfo{person}{Donald~W. Schwendeman}.} \bibinfo{year}{2008}\natexlab{}.
\newblock \showarticletitle{Parallel Computation of Three-Dimensional Flows
  Using Overlapping Grids With Adaptive Mesh Refinement}.
\newblock \bibinfo{journal}{\emph{J. Comput. Phys.}} \bibinfo{volume}{227},
  \bibinfo{number}{16} (\bibinfo{year}{2008}), \bibinfo{pages}{7469--7502}.
\newblock
\showISSN{0021-9991}
\urldef\tempurl%
\url{https://doi.org/10.1016/j.jcp.2008.04.033}
\showDOI{\tempurl}


\bibitem[Horne and Mahesh(2019)]%
        {HorneMahesh19}
\bibfield{author}{\bibinfo{person}{Wyatt~J. Horne} {and}
  \bibinfo{person}{Krishnan Mahesh}.} \bibinfo{year}{2019}\natexlab{}.
\newblock \showarticletitle{A Massively-Parallel, Unstructured Overset Method
  for Mesh Connectivity}.
\newblock \bibinfo{journal}{\emph{J. Comput. Phys.}}  \bibinfo{volume}{376}
  (\bibinfo{year}{2019}), \bibinfo{pages}{585--596}.
\newblock
\showISSN{0021-9991}
\urldef\tempurl%
\url{https://doi.org/10.1016/j.jcp.2018.09.053}
\showDOI{\tempurl}


\bibitem[Isaac et~al\mbox{.}(2012)]%
        {IsaacBursteddeGhattas12}
\bibfield{author}{\bibinfo{person}{Tobin Isaac}, \bibinfo{person}{Carsten
  Burstedde}, {and} \bibinfo{person}{Omar Ghattas}.}
  \bibinfo{year}{2012}\natexlab{}.
\newblock \showarticletitle{Low-Cost Parallel Algorithms for 2:1 Octree
  Balance}. In \bibinfo{booktitle}{\emph{Proceedings of the 26th IEEE
  International Parallel {\&} Distributed Processing Symposium}},
  Vol.~\bibinfo{volume}{1}. \bibinfo{publisher}{IEEE}, \bibinfo{address}{1730
  Massachusetts Avenue, NW Washington, DC, United States},
  \bibinfo{pages}{426--437}.
\newblock
\urldef\tempurl%
\url{https://doi.org/10.1109/IPDPS.2012.47}
\showDOI{\tempurl}


\bibitem[Isaac et~al\mbox{.}(2015)]%
        {IsaacBursteddeWilcoxEtAl15}
\bibfield{author}{\bibinfo{person}{Tobin Isaac}, \bibinfo{person}{Carsten
  Burstedde}, \bibinfo{person}{Lucas~C. Wilcox}, {and} \bibinfo{person}{Omar
  Ghattas}.} \bibinfo{year}{2015}\natexlab{}.
\newblock \showarticletitle{Recursive algorithms for distributed forests of
  octrees}.
\newblock \bibinfo{journal}{\emph{SIAM Journal on Scientific Computing}}
  \bibinfo{volume}{37}, \bibinfo{number}{5} (\bibinfo{year}{2015}),
  \bibinfo{pages}{C497--C531}.
\newblock
\urldef\tempurl%
\url{https://doi.org/10.1137/140970963}
\showDOI{\tempurl}
\showeprint[arxiv]{1406.0089}~[cs.DC]


\bibitem[Jarkowski et~al\mbox{.}(2014)]%
        {JarkowskiWoodgateBarakosEtAl13}
\bibfield{author}{\bibinfo{person}{Mikolay Jarkowski}, \bibinfo{person}{Mark~A.
  Woodgate}, \bibinfo{person}{George~N. Barakos}, {and} \bibinfo{person}{Jacek
  Rokicki}.} \bibinfo{year}{2014}\natexlab{}.
\newblock \showarticletitle{Towards Consistent Hybrid Overset Mesh Methods for
  Rotorcraft {CFD}}.
\newblock \bibinfo{journal}{\emph{International Journal for Numerical Methods
  in Fluids}} \bibinfo{volume}{74}, \bibinfo{number}{8} (\bibinfo{year}{2014}),
  \bibinfo{pages}{543--576}.
\newblock
\urldef\tempurl%
\url{https://doi.org/10.1002/fld.3861}
\showDOI{\tempurl}
\showeprint{https://onlinelibrary.wiley.com/doi/pdf/10.1002/fld.3861}


\bibitem[Jude et~al\mbox{.}(2021)]%
        {JudeSitaramanWissink21}
\bibfield{author}{\bibinfo{person}{Dylan Jude}, \bibinfo{person}{Jay
  Sitaraman}, {and} \bibinfo{person}{Andrew~M. Wissink}.}
  \bibinfo{year}{2021}\natexlab{}.
\newblock \showarticletitle{An Octree-Based, {C}artesian {CFD} Solver for
  Helios on {CPU} and {GPU} Architectures.}. In \bibinfo{booktitle}{\emph{AIAA
  Scitech 2021 Forum}}. \bibinfo{publisher}{AIAA}, \bibinfo{address}{Reston,
  Virginia}, \bibinfo{pages}{0841}.
\newblock
\urldef\tempurl%
\url{https://doi.org/10.2514/6.2021-0841}
\showDOI{\tempurl}
\showeprint{https://arc.aiaa.org/doi/pdf/10.2514/6.2021-0841}


\bibitem[Kirby et~al\mbox{.}(2019)]%
        {KirbyBrazellYangEtAl19}
\bibfield{author}{\bibinfo{person}{Andrew~C. Kirby},
  \bibinfo{person}{Michael~J. Brazell}, \bibinfo{person}{Zhi Yang},
  \bibinfo{person}{Rajib Roy}, \bibinfo{person}{Behzad~R. Ahrabi},
  \bibinfo{person}{Michael~K. Stoellinger}, \bibinfo{person}{Jay Sitaraman},
  {and} \bibinfo{person}{Dimitri~J. Mavriplis}.}
  \bibinfo{year}{2019}\natexlab{}.
\newblock \showarticletitle{Wind Farm Simulations using an Overset
  {$hp$}-adaptive Approach with blade-resolved Turbine Models}.
\newblock \bibinfo{journal}{\emph{The International Journal of High Performance
  Computing Applications}} \bibinfo{volume}{33}, \bibinfo{number}{5}
  (\bibinfo{year}{2019}), \bibinfo{pages}{897--923}.
\newblock
\urldef\tempurl%
\url{https://doi.org/10.1177/1094342019832960}
\showDOI{\tempurl}


\bibitem[Meakin(1991)]%
        {Meakin91}
\bibfield{author}{\bibinfo{person}{Robert~L. Meakin}.}
  \bibinfo{year}{1991}\natexlab{}.
\newblock \showarticletitle{A New Method for Establishing Intergrid
  Communication Among Systems of Overset Grids}. In
  \bibinfo{booktitle}{\emph{10th Computational Fluid Dynamics Conference}}.
  \bibinfo{publisher}{author}, \bibinfo{address}{Honolulu},
  \bibinfo{pages}{1586}.
\newblock
\urldef\tempurl%
\url{https://doi.org/10.2514/6.1991-1586}
\showDOI{\tempurl}
\showeprint{https://arc.aiaa.org/doi/pdf/10.2514/6.1991-1586}


\bibitem[Morton(1966)]%
        {Morton66}
\bibfield{author}{\bibinfo{person}{Guy~M. Morton}.}
  \bibinfo{year}{1966}\natexlab{}.
\newblock \bibinfo{booktitle}{\emph{A computer Oriented Geodetic Data Base; and
  a New Technique in File Sequencing}}.
\newblock \bibinfo{type}{{T}echnical {R}eport}. \bibinfo{institution}{IBM Ltd.}
\newblock


\bibitem[Noack et~al\mbox{.}(2009)]%
        {NoackBogerKunzEtAl09}
\bibfield{author}{\bibinfo{person}{Ralph Noack}, \bibinfo{person}{David Boger},
  \bibinfo{person}{Robert Kunz}, {and} \bibinfo{person}{Pablo Carrica}.}
  \bibinfo{year}{2009}\natexlab{}.
\newblock \showarticletitle{{Suggar++}: {A}n Improved General Overset Grid
  Assembly Capability}. In \bibinfo{booktitle}{\emph{19th AIAA Computational
  Fluid Dynamics}}. \bibinfo{publisher}{AIAA}, \bibinfo{address}{San Antonio,
  Texas}, \bibinfo{pages}{3992}.
\newblock
\urldef\tempurl%
\url{https://doi.org/10.2514/6.2009-3992}
\showDOI{\tempurl}


\bibitem[Noack and Kadanthot(2002)]%
        {NoackKadanthot02}
\bibfield{author}{\bibinfo{person}{RW Noack} {and} \bibinfo{person}{T
  Kadanthot}.} \bibinfo{year}{2002}\natexlab{}.
\newblock \showarticletitle{An octree based overset grid hole cutting method}.
  In \bibinfo{booktitle}{\emph{Proceedings of 8th International Conference On
  Numerical Grid Generation in Computational Field Simulations, Honolulu, HI}}.
  \bibinfo{publisher}{{NSF} Engineering Research Center of Computational Field
  Simulation}, \bibinfo{address}{Mississippi, USA}, \bibinfo{pages}{783--792}.
\newblock


\bibitem[Péron and Benoit(2013)]%
        {PeronBenoit13}
\bibfield{author}{\bibinfo{person}{Stéphanie Péron} {and}
  \bibinfo{person}{Christophe Benoit}.} \bibinfo{year}{2013}\natexlab{}.
\newblock \showarticletitle{Automatic Off-Body Overset Adaptive {C}artesian
  Mesh Method Based on an Octree Approach}.
\newblock \bibinfo{journal}{\emph{J. Comput. Phys.}} \bibinfo{volume}{232},
  \bibinfo{number}{1} (\bibinfo{year}{2013}), \bibinfo{pages}{153--173}.
\newblock
\showISSN{0021-9991}
\urldef\tempurl%
\url{https://doi.org/10.1016/j.jcp.2012.07.029}
\showDOI{\tempurl}


\bibitem[Roget and Sitaraman(2014)]%
        {RogetSitaraman14}
\bibfield{author}{\bibinfo{person}{Beatrice Roget} {and}
  \bibinfo{person}{Jayanarayanan Sitaraman}.} \bibinfo{year}{2014}\natexlab{}.
\newblock \showarticletitle{Robust and efficient overset grid assembly for
  partitioned unstructured meshes}.
\newblock \bibinfo{journal}{\emph{J. Comput. Phys.}}  \bibinfo{volume}{260}
  (\bibinfo{year}{2014}), \bibinfo{pages}{1--24}.
\newblock
\showISSN{0021-9991}
\urldef\tempurl%
\url{https://doi.org/10.1016/j.jcp.2013.12.021}
\showDOI{\tempurl}


\bibitem[Schluter et~al\mbox{.}(2005)]%
        {SchluterWuVanDerWeideEtAl05}
\bibfield{author}{\bibinfo{person}{Jorg Schluter}, \bibinfo{person}{Xiaohua
  Wu}, \bibinfo{person}{Edwin van~der Weide}, \bibinfo{person}{Seonghyeon
  Hahn}, \bibinfo{person}{Juan Alonso}, {and} \bibinfo{person}{Heinz Pitsch}.}
  \bibinfo{year}{2005}\natexlab{}.
\newblock \showarticletitle{Multi-Code Simulations: {A} Generalized Coupling
  Approach}. In \bibinfo{booktitle}{\emph{17th AIAA Computational Fluid
  Dynamics Conference}}. \bibinfo{publisher}{AIAA}, \bibinfo{address}{Toronto,
  Ontario, Canada}, \bibinfo{pages}{4997}.
\newblock
\urldef\tempurl%
\url{https://doi.org/10.2514/6.2005-4997}
\showDOI{\tempurl}
\showeprint{https://arc.aiaa.org/doi/pdf/10.2514/6.2005-4997}


\bibitem[Schwarz(2005)]%
        {Schwarz05}
\bibfield{author}{\bibinfo{person}{Thorsten Schwarz}.}
  \bibinfo{year}{2005}\natexlab{}.
\newblock \bibinfo{booktitle}{\emph{Ein blockstrukturiertes Verfahren zur
  Simulation der Umstr{\"o}mung komplexer Konfigurationen}}.
\newblock \bibinfo{type}{{T}echnical {R}eport}. \bibinfo{institution}{TU
  Braunschweig}.
\newblock
\urldef\tempurl%
\url{https://elib.dlr.de/44080/}
\showURL{%
\tempurl}


\bibitem[Shibliyev and Sezai(2021)]%
        {ShibilyevSezai21}
\bibfield{author}{\bibinfo{person}{Orxan Shibliyev} {and}
  \bibinfo{person}{Ibrahim Sezai}.} \bibinfo{year}{2021}\natexlab{}.
\newblock \showarticletitle{Overset Grid Assembler and Flow Solver with
  Adaptive Spatial Load Balancing}.
\newblock \bibinfo{journal}{\emph{Applied Sciences}}  \bibinfo{volume}{11}
  (\bibinfo{year}{2021}), \bibinfo{pages}{no. 5132}.
\newblock
\showISSN{2076-3417}
\urldef\tempurl%
\url{https://doi.org/10.3390/app11115132}
\showDOI{\tempurl}


\bibitem[Snively(2013)]%
        {Snively13}
\bibfield{author}{\bibinfo{person}{Jonathan~B. Snively}.}
  \bibinfo{year}{2013}\natexlab{}.
\newblock \showarticletitle{Mesospheric Hydroxyl Airglow Signatures of Acoustic
  and Gravity Waves Generated by Transient Tropospheric Forcing}.
\newblock \bibinfo{journal}{\emph{Geophysical Research Letters}}
  \bibinfo{volume}{40}, \bibinfo{number}{17} (\bibinfo{year}{2013}),
  \bibinfo{pages}{4533--4537}.
\newblock
\urldef\tempurl%
\url{https://doi.org/10.1002/grl.50886}
\showDOI{\tempurl}
\showeprint{https://agupubs.onlinelibrary.wiley.com/doi/pdf/10.1002/grl.50886}


\bibitem[Snively et~al\mbox{.}(2024a)]%
        {SnivelyCalhounAitonEtAl24a}
\bibfield{author}{\bibinfo{person}{Jonathan~B Snively}, \bibinfo{person}{Donna
  Calhoun}, \bibinfo{person}{Scott Aiton}, \bibinfo{person}{Carsten Burstedde},
  \bibinfo{person}{Hannes Brandt}, \bibinfo{person}{Tim Griesbach},
  \bibinfo{person}{Benedict Pineyro}, \bibinfo{person}{Pavel Inchin},
  \bibinfo{person}{Roberto Sabatini}, \bibinfo{person}{Michael Hirsch}, {and}
  \bibinfo{person}{Matthew~D Zettergren}.} \bibinfo{year}{2024}\natexlab{a}.
\newblock \bibinfo{title}{Scalable Modeling of Infrasound Propagation from
  Transient Sources in Surface-to-Exobase Domains}.
\newblock \bibinfo{howpublished}{AGU24 Annual Meeting}.
\newblock


\bibitem[Snively et~al\mbox{.}(2025a)]%
        {SnivelyCalhounAitonEtAl25a}
\bibfield{author}{\bibinfo{person}{Jonathan~B Snively}, \bibinfo{person}{Donna
  Calhoun}, \bibinfo{person}{Scott Aiton}, \bibinfo{person}{Carsten Burstedde},
  \bibinfo{person}{Hannes Brandt}, \bibinfo{person}{Tim Griesbach},
  \bibinfo{person}{Benedict Pineyro}, \bibinfo{person}{Pavel Inchin},
  \bibinfo{person}{Roberto Sabatini}, \bibinfo{person}{Michael Hirsch}, {and}
  \bibinfo{person}{Matthew~D Zettergren}.} \bibinfo{year}{2025}\natexlab{a}.
\newblock \bibinfo{title}{Scalable Modeling of Infrasound Propagation in
  Realistic Environments from Surface to Exobase}.
\newblock \bibinfo{howpublished}{AGU25 Annual Meeting}.
\newblock


\bibitem[Snively et~al\mbox{.}(2024b)]%
        {SnivelyCalhounAitonEtAl24}
\bibfield{author}{\bibinfo{person}{Jonathan~B Snively}, \bibinfo{person}{Donna
  Calhoun}, \bibinfo{person}{Scott Aiton}, \bibinfo{person}{Carsten Burstedde},
  \bibinfo{person}{Hannes Brandt}, \bibinfo{person}{Tim Griesbach},
  \bibinfo{person}{Benedict Pineyro}, \bibinfo{person}{Sehin Mesfin},
  \bibinfo{person}{Jaime~Aguilar Guerrero}, \bibinfo{person}{Christopher~James
  Heale}, \bibinfo{person}{Pavel Inchin}, \bibinfo{person}{Roberto Sabatini},
  \bibinfo{person}{Michael Hirsch}, {and} \bibinfo{person}{Matthew~D
  Zettergren}.} \bibinfo{year}{2024}\natexlab{b}.
\newblock \bibinfo{title}{Scalable Modeling of Acoustic-Gravity Wave-driven
  Traveling Ionospheric Disturbances}.
\newblock \bibinfo{howpublished}{AGU24 Annual Meeting}.
\newblock


\bibitem[Snively et~al\mbox{.}(2025b)]%
        {SnivelyCalhounAitonEtAl25}
\bibfield{author}{\bibinfo{person}{Jonathan~B Snively}, \bibinfo{person}{Donna
  Calhoun}, \bibinfo{person}{Scott Aiton}, \bibinfo{person}{Carsten Burstedde},
  \bibinfo{person}{Hannes Brandt}, \bibinfo{person}{Tim Griesbach},
  \bibinfo{person}{Benedict Pineyro}, \bibinfo{person}{Sehin Mesfin},
  \bibinfo{person}{Jaime~Aguilar Guerrero}, \bibinfo{person}{Christopher~J
  Heale}, \bibinfo{person}{Pavel Inchin}, \bibinfo{person}{Roberto Sabatini},
  \bibinfo{person}{Michael Hirsch}, {and} \bibinfo{person}{Matthew~D
  Zettergren}.} \bibinfo{year}{2025}\natexlab{b}.
\newblock \bibinfo{title}{Scalable Modeling of Acoustic-Gravity Wave
  Propagation and Evolutions from Impulsive Sources}.
\newblock \bibinfo{howpublished}{AGU25 Annual Meeting}.
\newblock


\bibitem[Snively et~al\mbox{.}(2023)]%
        {SnivelyCalhounAitonEtAl23}
\bibfield{author}{\bibinfo{person}{Jonathan~B Snively}, \bibinfo{person}{Donna
  Calhoun}, \bibinfo{person}{Scott Aiton}, \bibinfo{person}{Carsten Burstedde},
  \bibinfo{person}{Hannes Brandt}, \bibinfo{person}{Michael Hirsch},
  \bibinfo{person}{Jaime~Aguilar Guerrero}, \bibinfo{person}{Christopher~James
  Heale}, \bibinfo{person}{Pavel Inchin}, \bibinfo{person}{Sehin Mesfin},
  \bibinfo{person}{Benedict Pinyero}, \bibinfo{person}{Roberto Sabatini}, {and}
  \bibinfo{person}{Matthew~D Zettergren}.} \bibinfo{year}{2023}\natexlab{}.
\newblock \bibinfo{title}{Scalable Modeling of Acoustic-Gravity Wave Processes
  in the Ionosphere, Thermosphere, and Mesosphere {(ITM)}}.
\newblock \bibinfo{howpublished}{AGU23 Fall Meeting}.
\newblock


\bibitem[Steger et~al\mbox{.}(1983)]%
        {StegerDoughertyBenek83}
\bibfield{author}{\bibinfo{person}{Joseph~L. Steger},
  \bibinfo{person}{F.~Carroll Dougherty}, {and} \bibinfo{person}{John~A.
  Benek}.} \bibinfo{year}{1983}\natexlab{}.
\newblock \showarticletitle{A Chimera grid scheme}.
\newblock \bibinfo{journal}{\emph{K.N. Ghia, U. Chia (Eds.), Advance in Grid
  Generation, ASME FED}}  \bibinfo{volume}{5} (\bibinfo{year}{1983}),
  \bibinfo{pages}{59 -- 69}.
\newblock


\bibitem[Suhs et~al\mbox{.}(2002)]%
        {SuhsRogersDietz02}
\bibfield{author}{\bibinfo{person}{Norman Suhs}, \bibinfo{person}{Stuart
  Rogers}, {and} \bibinfo{person}{William Dietz}.}
  \bibinfo{year}{2002}\natexlab{}.
\newblock \showarticletitle{{PEGASUS} 5: {A}n Automated Pre-Processor for
  Overset-Grid {CFD}}. In \bibinfo{booktitle}{\emph{32nd AIAA Fluid Dynamics
  Conference and Exhibit}}. \bibinfo{publisher}{AIAA}, \bibinfo{address}{St.
  Louis, Missouri}, \bibinfo{pages}{3186}.
\newblock
\urldef\tempurl%
\url{https://doi.org/10.2514/6.2002-3186}
\showDOI{\tempurl}
\showeprint{https://arc.aiaa.org/doi/pdf/10.2514/6.2002-3186}


\bibitem[Sundar et~al\mbox{.}(2008)]%
        {SundarSampathBiros08}
\bibfield{author}{\bibinfo{person}{Hari Sundar}, \bibinfo{person}{Rahul
  Sampath}, {and} \bibinfo{person}{George Biros}.}
  \bibinfo{year}{2008}\natexlab{}.
\newblock \showarticletitle{Bottom-up construction and 2:1 balance refinement
  of linear octrees in parallel}.
\newblock \bibinfo{journal}{\emph{SIAM Journal on Scientific Computing}}
  \bibinfo{volume}{30}, \bibinfo{number}{5} (\bibinfo{year}{2008}),
  \bibinfo{pages}{2675--2708}.
\newblock
\urldef\tempurl%
\url{https://doi.org/10.1137/070681727}
\showDOI{\tempurl}


\bibitem[Zettergren et~al\mbox{.}(2025)]%
        {ZettergrenCalhounSnivelyEtAl25}
\bibfield{author}{\bibinfo{person}{Matthew~D Zettergren},
  \bibinfo{person}{Donna Calhoun}, \bibinfo{person}{Jonathan~B Snively},
  \bibinfo{person}{Carsten Burstedde}, \bibinfo{person}{Hannes Brandt},
  \bibinfo{person}{Pavel Inchin}, \bibinfo{person}{Michael Hirsch},
  \bibinfo{person}{Tim Griesbach}, \bibinfo{person}{Scott Aiton}, {and}
  \bibinfo{person}{Kshitija Deshpande}.} \bibinfo{year}{2025}\natexlab{}.
\newblock \bibinfo{title}{Numerical Modeling of Ionospheric Responses to
  Ground-level, Anthropogenic Sources}.
\newblock \bibinfo{howpublished}{AGU25 Annual Meeting}.
\newblock


\bibitem[Zettergren and Snively(2015)]%
        {ZettergrenSnively15}
\bibfield{author}{\bibinfo{person}{Matthew~D. Zettergren} {and}
  \bibinfo{person}{Jonathan~B. Snively}.} \bibinfo{year}{2015}\natexlab{}.
\newblock \showarticletitle{Ionospheric response to infrasonic-acoustic waves
  generated by natural hazard events}.
\newblock \bibinfo{journal}{\emph{Journal of Geophysical Research: Space
  Physics}} \bibinfo{volume}{120}, \bibinfo{number}{9} (\bibinfo{year}{2015}),
  \bibinfo{pages}{8002--8024}.
\newblock
\urldef\tempurl%
\url{https://doi.org/10.1002/2015JA021116}
\showDOI{\tempurl}
\showeprint{https://agupubs.onlinelibrary.wiley.com/doi/pdf/10.1002/2015JA021116}


\bibitem[Zettergren and Snively(2019)]%
        {ZettergrenSnively19}
\bibfield{author}{\bibinfo{person}{Matthew~D. Zettergren} {and}
  \bibinfo{person}{Jonathan~B. Snively}.} \bibinfo{year}{2019}\natexlab{}.
\newblock \showarticletitle{Latitude and Longitude Dependence of Ionospheric
  {TEC} and Magnetic Perturbations From Infrasonic Acoustic Waves Generated by
  Strong Seismic Events}.
\newblock \bibinfo{journal}{\emph{Geophysical Research Letters}}
  \bibinfo{volume}{46}, \bibinfo{number}{3} (\bibinfo{year}{2019}),
  \bibinfo{pages}{1132--1140}.
\newblock
\urldef\tempurl%
\url{https://doi.org/10.1029/2018GL081569}
\showDOI{\tempurl}


\bibitem[Zhao et~al\mbox{.}(2024)]%
        {ZhaoLiGuoEtAl24}
\bibfield{author}{\bibinfo{person}{Ran Zhao}, \bibinfo{person}{Chao Li},
  \bibinfo{person}{Xiaowei Guo}, \bibinfo{person}{Sen Zhang},
  \bibinfo{person}{Xi Yang}, \bibinfo{person}{Tao Tang}, {and}
  \bibinfo{person}{Canqun Yang}.} \bibinfo{year}{2024}\natexlab{}.
\newblock \bibinfo{booktitle}{\emph{A Motion Trace Decomposition-based overset
  grid method for parallel {CFD} simulations with moving boundaries}}.
\newblock \bibinfo{publisher}{Association for Computing Machinery},
  \bibinfo{address}{New York, NY, USA}, \bibinfo{pages}{411–420}.
\newblock
\showISBNx{9798400717932}
\urldef\tempurl%
\url{https://doi.org/10.1145/3673038.3673102}
\showURL{%
\tempurl}


\end{thebibliography}

\appendix

\section{Application to Disjoint Communicators}\label{app:arc_disjoint_communicators}

In \secref{sec:applications} we only performed tests for congruent \cons{} and
\pro{} communicators with both of them being equal to the global communicator.
However, the overset algorithm can also be performed for disjoint or
partially overlapping communicators.
How the code is distributed between the communicators is shown in the
pseudocode of \algref{alg:overset}{3}.

We adapt our implementation to deal with disjoint or partially overlapping
communicators.
To achieve this, without loss of generality we define the \cons{} communicator
as the subset $[0,\Pcons)$ and the \pro{} communicator as the subset
$[\Ppro,P_\text{global})$ of \texttt{MPI\_COMM\_WORLD}, where
$P_\text{global}$ denotes the global process count.

The overset algorithm requires extensive point-to-point communication between
processes from the two communicators.
Since both communicators are contiguous subsets of \texttt{MPI\_COMM\_WORLD},
it suffices to exchange their respective global communicator offsets $0$ and
$\Ppro$.
Based on this information, the point-to-point communication between \cons{} and
\pro{} can be performed on the global communicator.

One of the major differences between the congruent and the disjoint case is
the communication of basic \pro{} mesh information such as the global first
positions $\arry F$, the amount of \pro{} trees and the size of the \pro{}
communicator.
If the communicators are congruent, or the \pro{} communicator completely
contains the \cons{} communicator, the \cons{} processes can access this
information directly.
In all other cases, the \pro{} root sends the information to the \cons{} root
for
broadcasting on the \cons{} communicator.
The length of $\arry F$ depends on the amount of \pro{} processes, so these
pieces of data need to be communicated in separate iterations.

We tested our implementation for the arc example as described in
\secref{sec:arc_simulation_setup} for $\Pcons = \Ppro =
\frac{P_\text{global}}{2}$.
The results are too similar to the strong scaling results for congruent
communicators (see \figref{fig:timings_arc_strong}) to justify the inclusion of
another plot.
In particular, the run time does not improve when switching from $\Ppro =
\Pcons$ processes in congruent communicators to the same amount of \pro{} and
\cons{} processes in disjoint communicators.
This is to be expected, since there is little potential for overlap between the
\cons{} and \pro{} tasks.
The \pro{} processes have to wait for the \cons{} processes to finish their
partition search and send the resulting query point buffers, before they can
start the local search and evaluation.
Vice versa, the \cons{} processes have to wait until the \pro{} processes return
the evaluation data to incorporate it into their local mesh.

So, the disjoint communicator approach does not yield any advantages for the
mesh overset itself.
However, it offers the possibility to overlap the mesh overset with other \pro{}
or \cons{} specific tasks, which can be performed while waiting for the
partition/local search results.
In particular, it is possible to switch to disjoint communicators without
losing scalability.

\section{Reproducing Examples}
The numerical applications presented in \secref{sec:applications} cover
different example settings implemented in the \texttt{example/multi} directory
of the \pforest{} library~\cite{BursteddeWilcoxIsaac25} under commit id
b34b7162.

The examples can be executed by calling
\begin{equation*}
  \texttt{mpirun -np \$\{n\} ./example/multi/p\{4,8\}est\_overlap}
\end{equation*}
with the following command line parameters

\begin{itemize}[labelindent=1em,labelsep=1cm,leftmargin=*,align=parleft]
	\item[\texttt{-e}] Specify the mappings for the examples.\\
		$0$ - Arc and brick mapping from \secref{sec:arc_simulation_setup} with
		      switched \cons{} and \pro{} roles.\\
		$1$ - Arc and brick mapping from \secref{sec:arc_simulation_setup}.\\
    $2$ - Rotated unit square mappings from \secref{sec:unit_square_example}.\\
    $3$ - Un-rotated unit square mappings from \secref{sec:unit_square_example}.
	\item[\texttt{-r}] Specify the refinement method.\\
	  $0$ - Refinement around the intersection area from
	        \secref{sec:adaptive_refinement}.\\
	  $1$ - Refinement around a specific point from
	        \secref{sec:arc_simulation_setup}.\\
	  $2$ - Artificial refinement based on child-id and rank of the octants.\\
	  $3$ - Refinement around a pentagon from
	        \ref{sec:unit_square_adaptive_refinement}.
	\item[\texttt{-l}] Specify if the load balancing from
	  \secref{sec:load_balancing} should be enabled.
	  Defaults to \texttt{false}.
	\item[\texttt{-a,-b}] Specify the \cons{} and \pro{} communicator sizes.
		The \cons{} communicator covers the range $[0,\texttt{a})$ of ranks.
		The \pro{} communicator covers the range $[\texttt{b},n)$.
		Both default to the range $[0,n)$.
	\item[\texttt{-c,-p}] Specify the initial uniform refinement level of \cons{}
	  and \pro{} mesh. Both default to $0$.
	\item[\texttt{-m}] Specify the maximum refinement level for the iterative
	  refinement.
	  Defaults to $0$.
	\item[\texttt{-v,-t}] Specify the form of the output.
		If \texttt{v} is \texttt{true}, vtk output is generated for both meshes.
		If \texttt{t} is \texttt{true}, a list of all query points with their
		evaluation data is printed to the console.
		Both default to \texttt{false}.
\end{itemize}

The scaling tests for the unit square example with adaptive refinement from
\ref{sec:unit_square_example} were performed calling
{
\setlength{\parindent}{0mm}

\bigskip
$\begin{aligned}
	\texttt{./example/multi/p4est\_overlap -c 2 -p 2 -r 3 -m 20 -e \{2,3\}}
\end{aligned}$\bigskip

with \texttt{e} being switched for rotated and unrotated versions.
The scaling tests for the arc example from \secref{sec:arc_numerical_results}
and \secref{sec:load_balancing} were performed calling

\bigskip
$\begin{aligned}
	\texttt{./example/multi/p8est\_overlap -c \{5,6\} -p \{4,5\} -r 1
		-m \{9,10\} -e 1 -l \{0,1\}}
\end{aligned}$\bigskip

with \texttt{c}, \texttt{p} and \texttt{m} being switched for the small and
middle example and \texttt{l} being switched for enabling or disabling load
balancing.
Similarly, the weak scaling tests were performed calling:

\bigskip
$\begin{aligned}
	\texttt{./example/multi/p8est\_overlap } &\texttt{-c \{4,5,6,7\}
		-p \{3,4,5,6\} -r 1 -m \{8,9,10,11\}}\\
		&\texttt{-e 1 -l 1}
\end{aligned}$\bigskip

Finally, the scaling tests for disjoint communicators from
\appref{app:arc_disjoint_communicators} were performed calling:

\bigskip
$\begin{aligned}
	\texttt{./example/multi/p8est\_overlap } &\texttt{-c \{5,6\} -p \{4,5\} -r 1
		-m \{9,10\} -e 1}\\
	 &\texttt{-a n/2 -b n/2}
\end{aligned}$\bigskip
}

\end{document}